\documentclass[twocolumn, trackchanges]{aastex62}

\usepackage[latin1]{inputenc}
\usepackage{amsmath}
\usepackage{wasysym}
\usepackage{amsfonts}
\usepackage{amssymb}
\usepackage{graphicx}
\usepackage{listings}
\usepackage{verbatim}
\usepackage{natbib}
\usepackage{hyperref}
\begin{abstract}
We investigate the ratio of coronal and transition region intensity in coronal loops observed by the Atmospheric Imaging Assembly (AIA) on the Solar Dynamics Observatory (SDO). Using Enthalpy-based Thermal Evolution of Loops (EBTEL) hydrodynamic simulations, we model loops with multiple lengths and energy fluxes heated randomly by events drawn from power-law distributions with different slopes and minimum delays between events to investigate how each of these parameters influences observable loop properties. We generate AIA intensities from the corona and transition region for each realization. The variations within and between models generated with these different parameters illustrate the sensitivity of narrowband imaging to the details of coronal heating. We then analyze the transition region and coronal emission from a number of observed active regions and find broad agreement with the trends in the models. In both models and observations, the transition region brightness is significant, often greater than the coronal brightness in all six ``coronal'' AIA channels. We also identify an inverse relationship, consistent with heating theories, between the slope of the differential emission measure (DEM) coolward of the peak temperature and the observed ratio of coronal to transition region intensity. These results highlight the use of narrowband observations and the importance of properly considering the transition region in investigations of coronal heating.
\end{abstract}

\begin{document}
\title{Transition region contribution to AIA observations in the context of coronal heating}
\author{S. J. Schonfeld}
\affiliation{Institute for Scientific Research, Boston College, Newton, MA 02459, schonfsj@gmail.com}
\author{J. A. Klimchuk}
\affiliation{NASA Goddard Space Flight Center, Heliophysics Science Division, Greenbelt, MD 20771}
\date{\today}

\section{Introduction}
A consensus understanding of how exactly the plasma of the Sun's corona is heated to MK temperatures has remained elusive for decades \citep[for more details see reviews by:][]{Zirker1993, Walsh2003, Klimchuk2006, Klimchuk2015, Parnell2012, Viall2020}. Many physical mechanisms have been proposed to cause this heating \citep[for lists of many such mechanisms see:][]{Mandrini2000, Cranmer2019}, but the observations needed to distinguish them are fundamentally challenging. The basic difficulty is that, for all mechanisms, the heating is highly time dependent with a small (generally subresolution) spatial scale perpendicular to the magnetic field. In this context, it is convenient to consider magnetic strands, bundles of magnetic flux with approximately uniform plasma properties over their cross section \citep{Klimchuk2006}. These properties evolve in time in a manner that depends strongly on the details of the heating in the strand. The optically thin nature of coronal plasma emission in the extreme ultraviolet (EUV) and X-ray results in confusion between the many overlapping strands along a line of sight \citep[e.g.;][]{Viall2011}. This makes it impossible to study the dynamics of a single heating event in isolation.

Instead, coronal heating must be studied by determining how the bulk, optically thin plasma responds to heating on observable scales \citep{Hinode2019}. By simulating the observable response of plasma to heating on unobservably small scales it is possible to constrain the properties of the heating with available instrumentation. This is commonly done by simulating the evolution of plasma within individual closed magnetic strands \citep[e.g.;][]{Barnes2016a, Barnes2016b} and then generating the emission due to collections of these strands \citep{Cargill1994, Patsourakos2008, Warren2002, Cargill2004, Warren2010a, Bradshaw2011, Reep2013, Viall2013, Lionello2016, Marsh2018} in observable instrument channels.

These coronal models must necessarily consider the coupled system with the transition region that moderates the connection between the hot, tenuous corona and the cool, dense chromosphere. In observational terms, the transition region has commonly been defined based on the temperature regime it occupies, $\sim10^4$ -- $10^6$ K. A more appropriate and physically motivated definition is given by considering models of individual magnetic strands and defining the interface between the corona and transition region to be the location where thermal conduction changes from being a loss term above (causing cooling) to a gain term below \citep[causing heating;][]{Vesecky1979}. This is the approach taken in the Enthalpy-based Thermal Evolution of Loops (EBTEL) model originally defined in \cite{Klimchuk2008}. The advantage of this definition is that it more faithfully represents the range of possible states available to coronal loop transition regions. In particular, this acknowledges that in hot loops, temperatures commonly associated with the corona (above $10^6$ K) can occur in the transition region close to the loop footpoints where the density and temperature gradients are large. It also allows for the transition region of an individual loop to evolve dynamically in time in response to the heating and cooling of the loop as a whole \citep{Johnston2017a, Johnston2017b, Johnston2019}.

Despite being a small fraction of the volume of a loop \citep[both because it is confined to near the footpoints and because the cross-sectional area of a loop typically increases substantially between the high-$\beta$ photosphere and the low-$\beta$ corona, e.g.;][]{Guarrasi2014} the higher densities in the transition region mean that it emits brightly in the EUV. Therefore, the origin of observed EUV emission from lines that emit in the few MK range is not \textit{a priori} clear. This emission could originate from relatively cooler coronal loops or from the transition regions of much hotter loops. This uncertainty is the motivation for the present study, to determine how much coronal and transition region emission is expected from loop models in the various Atmospheric Imaging Assembly \citep[AIA;][]{Lemen2012} channels and how this compares with observations. In Section \ref{sec:simulation} we briefly describe the EBTEL model and the results of varying loop and heating parameters on the modeled AIA emission. In Section \ref{sec:observations} we develop a simple procedure to estimate the transition region contribution in AIA observations and apply it to a number of active regions. We summarize our findings and comment on the implications of these results in Section \ref{sec:conclusion}.

\section{EBTEL hydrodynamic simulations}
\label{sec:simulation}
EBTEL \citep[``enthalpy-based thermal evolution of loops'';][]{Klimchuk2008, Cargill2012a, Cargill2012b} models the time evolution of the coronal-averaged temperature, pressure, and density in a single magnetic strand in 0D. It is able to accurately describe subsonic plasma evolution under gentle and impulsive heating and can approximately treat complex phenomena such as saturated heat flux and nonthermal electron beam heating. A significant feature of EBTEL is its speed; it can compute the evolution of a single magnetic strand for one day of physical time in seconds, orders of magnitude faster than comparable 1D models. Despite the simplicity of the model, EBTEL's results are very similar to the spatial average determined along the length of a 1D simulation \citep{Klimchuk2008, Cargill2012a, Cargill2012b}. In addition to computing the average coronal properties, EBTEL also determines the coronal and transition region Differential Emission Measures (DEMs) at each time step. Here we use the EBTEL++ implementation described in \citep{Barnes2016a} and available online at \url{https://github.com/rice-solar-physics/ebtelPlusPlus}.

One of the simplifications necessary in the formulation of EBTEL is an assumed ratio of the radiative losses in the transition region and corona. In the model, this is represented by the semiconstant $c_{1}=R_{tr}/R_{c}$ where $R_{tr}$ and $R_{c}$ are the total radiative losses from the transition region and corona, respectively. This ratio depends on the fixed length of the strand, the dynamic plasma temperature (which influences the plasma scale height), and the coronal density ($n$) relative to the static equilibrium density for a loop with the same temperature ($n_{eq}$). At low coronal densities ($n < n_{eq}$) conduction dominates the coronal losses and the relative transition region emission is particularly strong \citep{Barnes2016a}. At high densities ($n > n_{eq}$) coronal losses are dominated by radiation, and therefore the relative emission from the transition region is reduced \citep{Cargill2012a}. Ignoring corrections for gravitational stratification and details of the radiative loss function, which are included in EBTEL, this ratio smoothly varies with density between the limits 
\begin{equation}
c_{1}=\frac{R_{tr}}{R_{c}} = \begin{cases}
								  2 & n \le n_{eq}   \\
								0.6 & n \gg n_{eq}
							 \end{cases}
\label{eqn:c1}	
\end{equation}
which have been chosen to produce results consistent with HYDrodynamic and RADiative emission (HYDRAD) 1D loop models \citep{Bradshaw2003a, Bradshaw2003b, Bradshaw2004, Bradshaw2013} for a wide range of coronal heating scenarios. It is important to note that while these prescriptions have a controlling influence on the total transition region and coronal emission, they do not directly impact the intensity of the individual channels investigated in this study. This is due to the nonuniform temperature response of the AIA channels (discussed in Section \ref{sec:simulation:AIA}) which results in their preferential sensitivity to plasma of particular temperatures. In a given (real or simulated) observation, a particular channel may measure emission from the transition region, corona, or a combination of the two, independent of $c_{1}$.

EBTEL defines two other constants related to the temperature profile of a 1D strand. One relates the average coronal temperature in the strand, $\bar T$, to the apex temperature, $T_{a}$:
\begin{equation}
c_{2}=\frac{\bar T}{T_{a}}=0.9
\label{eqn:c2}	
\end{equation}
and the other the temperature at the top of the transition region, $T_{0}$, to the apex temperature:
\begin{equation}
c_{3}=\frac{T_{0}}{T_{a}}=0.6.
\label{eqn:c3}	
\end{equation}
These values were chosen based on hydrostatic 1D models computed with HYDRAD, but they are found to be reasonable representations when subsonic flows are present. Equation \ref{eqn:c3} is particularly important for the current investigation since we are interested in the distinction between the transition region and corona. This means that the calculated coronal temperature determines the maximum temperature of the transition region, which is assumed to cover all temperatures between $T_{0}$ and chromospheric temperatures. We stress that $T_{0}$ is a physically motivated temperature that correctly demarcates the region of steep temperature and density gradients at the base of a coronal loop.

\subsection{Power-law distribution of heating events}
\label{sec:simulation:power_law}
\begin{figure}[t]
	\includegraphics*[trim=.1cm .1cm 0.85cm .25cm, clip, width=\columnwidth]
{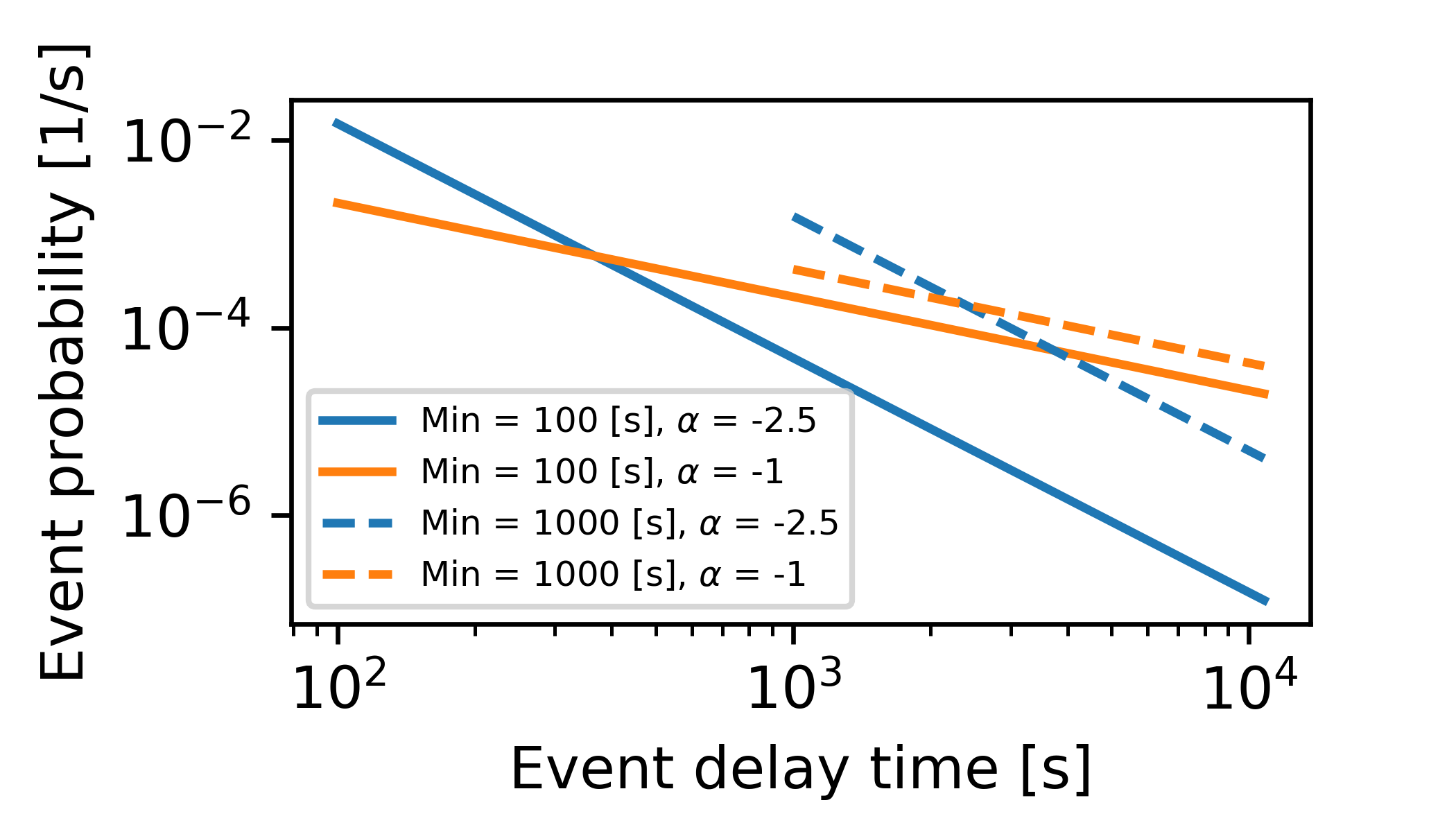}
	\caption{Power-law distributions of heating event delay times. The maximum of each distribution is 10,800 s (3 hr).}
	\label{fig:powerlaws}
\end{figure}
In these simulations we heat the plasma with a combination of a constant background heating ($1\%$ of the total energy input) and symmetrical triangular impulsive heating events. Each heating event has a duration of t$_{e}=100\text{s}$ and a total energy input per unit volume $\epsilon_{e}$ proportional to the delay time to the next event given by
\begin{equation}
\epsilon_{e}=0.99\ \text{t}_{d}\left(\frac{\text{F}}{\text{L}}\right)=0.5\ \text{Q}_{max}\text{t}_{e}
\label{eqn:event_energy}
\end{equation}
where t$_{d}$ is the random delay until the next event, $\text{Q}_{max}$ is the maximum volumetric heating rate during the event, and F and L are the energy flux and strand half length (in centimeters) given in table \ref{tab:parameters}. The factor of 0.99 accounts for the 1\% constant background heating. The result of this scaling is that each heating cycle has the same time-averaged volumetric heating rate, which is prescribed assuming that the deposited energy is evenly distributed over the length of the loop. The individual heating events are randomly drawn from power-law distributions of heating event delay time (t$_{d}$) shown in Figure \ref{fig:powerlaws}. These power laws are defined by their exponent ($\alpha$, the slope when visualized in log-log space) and minimum and maximum time delay between events. For all models, the maximum delay time is fixed at three hours (10,800 s), that is, each modeled magnetic strand experiences an impulsive heating event at least once every three hours.

This numerical scheme represents a physical system that is driven with a constant energy buildup rate that releases some fraction of this energy when a critical threshold value is reached. This is consistent with, for example, critical stress reconnection heating driven by random-walk footpoint motion \citep{Parker1988, Lopez_Fuentes2015}. In this mechanism, the stress in the magnetic field builds with time until a critical level defined in terms of the angle between adjacent magnetic strands is reached, at which point they reconnect and release a fraction of the energy stored in the field. The more energy that is released, the longer it will take for the magnetic field to return to the critical stressed state and reconnect again. Note, however, that the prescribed heating scheme used here does not assume any particular physical mechanism and is consistent with any heating scenario that builds to a threshold level. It also yields similar although not identical (due to the fact that the effects of a heating event are dependent on the physical state of the loop when heating begins) average conditions to systems with constant driving that build to a random stressed state before relaxing impulsively to some constant minimum energy state \citep{Cargill2014}. Similarly, it will emulate any system with a power-law distribution of heating event amplitudes and delay times.

\subsection{Modeled parameters}
\label{sec:simulation:parameters}
\begin{deluxetable}{lccc}
  \tablewidth{0pt}
  \tablecaption{EBTEL Model parameters}
  \tablecolumns{3}
  \tablehead{Parameter & \colhead{Symbol} & \colhead{Low Value} & \colhead{High Value}}
  \startdata
  	Strand half length [Mm] & L & $20$ & $80$\\
	Energy flux [$\text{erg cm}^{-2} \text{ s}^{-1}$] & F & $5\times10^{6}$ & $2\times10^{7}$\\
	Minimum delay [s] & t$_\text{min}$ & $100$ & $1000$\\
	Power-law slope & $\alpha$ & $-2.5$ & $-1.0$\\
  \enddata
  \tablecomments{Parameters of models changed for the different simulations. The parameters held constant in all runs are given in table \ref{tab:fixed}.\label{tab:parameters}}
\end{deluxetable}

We perform a parameter space exploration over relevant physical properties of coronal heating. This involves computing EBTEL hydrodynamic models for combinations of four parameters each in two different states for a total of 16 different conditions. These parameters are: the length of the magnetic strand, the time-averaged energy flux into the base of the strand (related to the time-averaged volumetric heating rate: $\text{Q}=\text{F}/\text{L}$), the minimum event delay time, and the power-law slope of the distribution of delay times. The parameters explored here represent typical (and by no means extreme) ranges for coronal active regions, where known. These parameters are listed in table \ref{tab:parameters} and described below.

\subsubsection{Strand length}
\label{sec:simulation:parameters:strand_length}
We simulate strands with half lengths (footpoint to apex) of 20 and 80 Mm, sizes typical of observable loops in coronal active regions (e.g. those examined in Section \ref{sec:observations}).

\subsubsection{Energy flux}
\label{sec:simulation:parameters:energy_flux}
The total energy losses from the corona in active regions (i.e. the heating necessary for consistency with observations) are $\sim 10^7\ \text{erg}\ \text{cm}^{-2}\ \text{s}^{-1}$ \citep{Withbroe1977} and we heat our models with half and twice this value to simulate weakly and strongly heated regions.

\subsubsection{Minimum delay between heating events}
\label{sec:simulation:parameters:delay}
``Time lag'' analysis of active regions using AIA observations suggests that the characteristic delay between successive heating events is similar to the plasma cooling timescale \citep{Viall2017}, which depends strongly on the loop length, but is on the order of a thousand seconds. On the other hand, theoretical considerations of reconnection-based heating suggest delays on the order of a hundred seconds \citep{Klimchuk2015}. We therefore test distributions with minimum delay times of 100 and 1000 s.

\subsubsection{Power-law slopes of event delays}
\label{sec:simulation:parameters:power_law}
Many observational studies suggest that flares occur with a power-law distribution \citep[e.g. see discussion in;][]{Parnell2012}, and power-law distributions of nanoflares can explain the observed range in DEM slopes coolward of the emission measure peak found in active regions \citep{Cargill2014}. Many theoretical models have also suggested that nanoflares occur with a power-law distribution in energy, from a simple cellular automaton \citep{Lopez_Fuentes2015} to full three-dimensional magnetohydrodynamic (MHD) simulations \citep{Knizhnik2018}. These models and observational considerations typically find nanoflare energy distributions with power laws of $-2.5 \lesssim \alpha \lesssim -1.5$. However, recent MHD simulations tracking discontinuities in field line tracing by \cite{Knizhnik2020} suggest nanoflares with time delay and energy power laws with  $\alpha \approx -1$. Consequently, our models test heating event power laws with $\alpha = -1$ and $\alpha = -2.5$. Due to the proportionality between the delay time and event energy (Section \ref{sec:simulation:power_law}), the energy input from the power laws with $\alpha = -2.5$ is small-event dominated while for $\alpha = -1$ it is evenly distributed between events smaller and larger than the average of the smallest and largest events.

\subsection{Model results}
\label{sec:simulation:setup}
\begin{deluxetable}{lc}
  \tablewidth{0pt}
  \tablecaption{EBTEL fixed model parameters}
  \tablecolumns{3}
  \tablehead{Keyword \textit{(description)} & \colhead{Value}}
  \startdata
  	total\_time \textit{(seconds)} & $10^{5}$\\
	tau \textit{(initial time step, seconds)} & $1.0$\\
	tau\_max \textit{(maximum time step, seconds)} & $50$\\
	force\_single\_fluid \textit{(electron-ion equilibrium)} & True\\
	use\_c1\_loss\_correction & True\\
	use\_c1\_grav\_correction & True\\
	use\_power\_law\_radiative\_losses & True\\
	use\_flux\_limiting \textit{(for conductive cooling)}& False\\
	use\_adaptive\_solver \textit{(for dynamic tau)} & True\\
	adaptive\_solver\_error & $1\times10^{-6}$\\
	adaptive\_solver\_safety & $0.5$\\
	c1\_cond \textit{(c1 during conductive cooling)} & $2.0$\\
	c1\_rad \textit{(c1 during radiative cooling}) & $0.6$\\
	helium\_to\_hydrogen\_ratio & $0.075$\\
	surface\_gravity \textit{(relative to solar)} & $1.0$\\
	dem\_use\_new\_method & True\\
	heating partition \textit{(1 = electron, 0 = ion)} & $0.5$\\
  \enddata
  \tablecomments{Relevant EBTEL parameters held constant for all simulations. More detailed descriptions of these keywords are provided through the EBTEL++ github repository at \url{https://rice-solar-physics.github.io/ebtelPlusPlus/configuration/}\label{tab:fixed}}
\end{deluxetable}

\begin{figure*}[t]
	\includegraphics[trim=0.45cm 0cm 7.5cm 0cm, clip, width=0.4925\textwidth]
{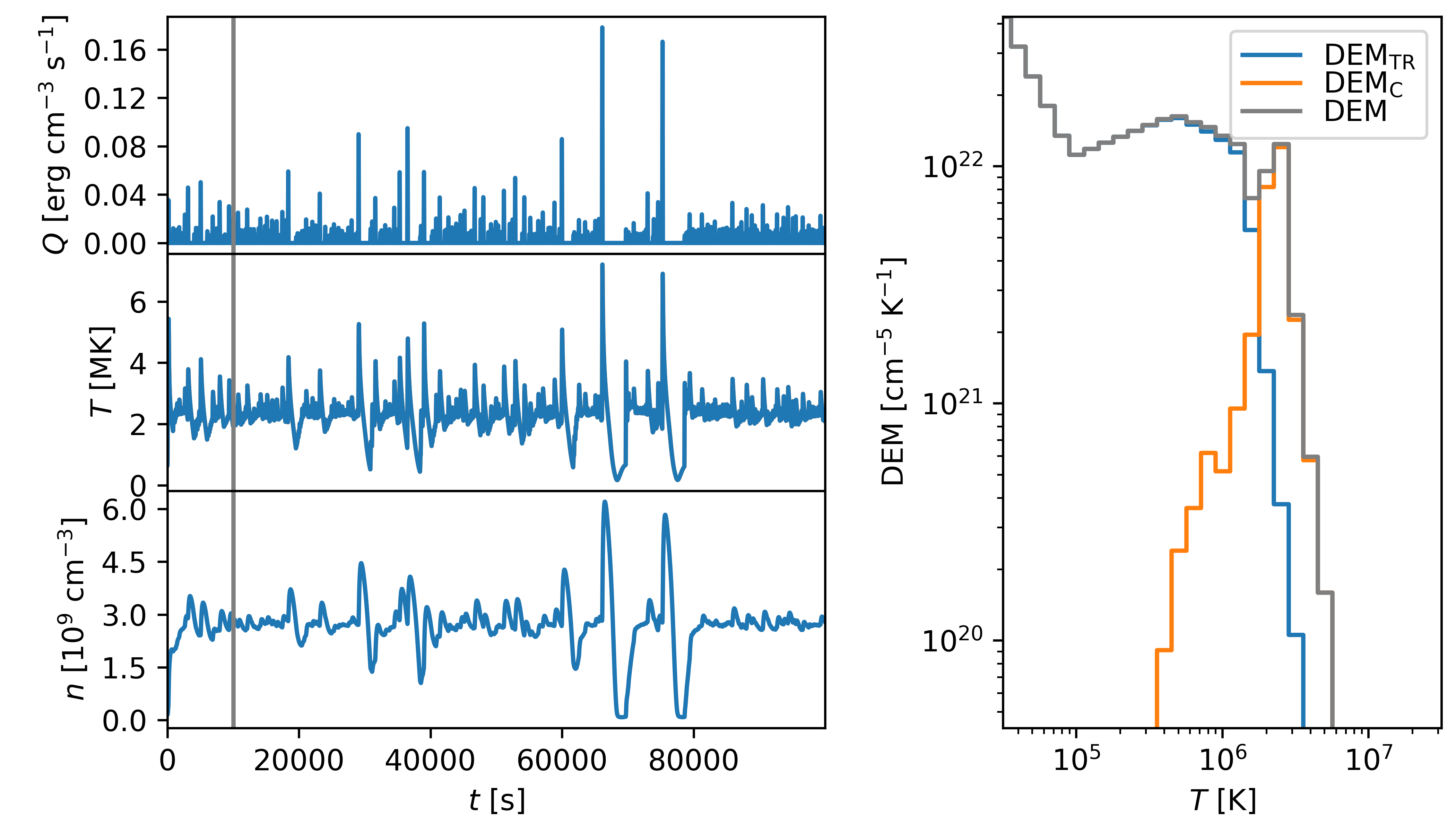}
	\hspace{0.1cm}
	\includegraphics[trim=0.45cm 0cm 7.5cm 0cm, clip, width=0.4925\textwidth]
{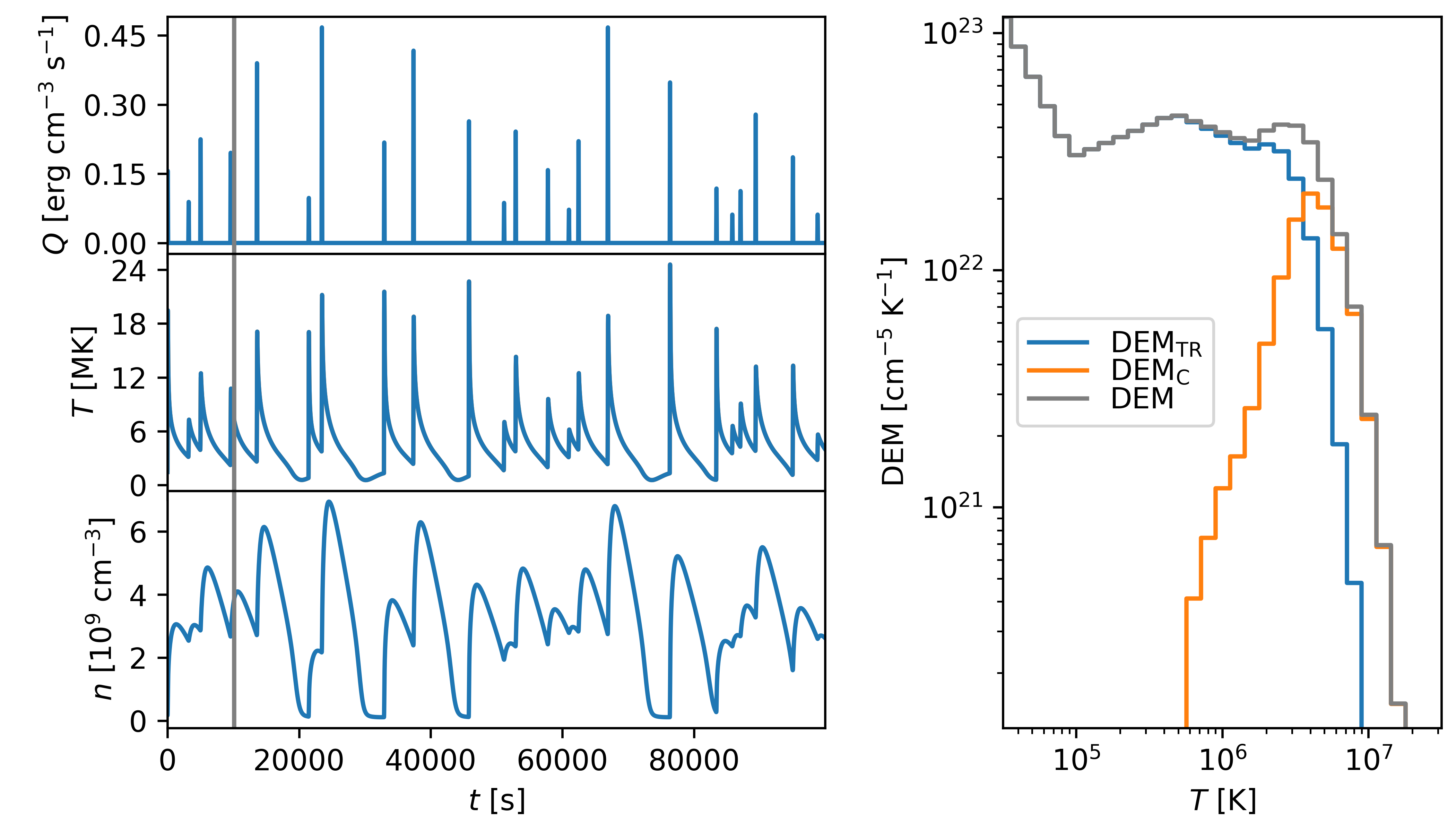}
	\caption{Time evolution of coronal parameters in EBTEL models of individual magnetic strands. \textit{Left}: a strand with a half length of $\text{L}=20$ Mm, average energy flux of $\text{F}=5\times10^6\ \text{erg}\ \text{cm}^{-2}\ \text{s}^{-1}$, minimum delay between events of t$_{\text{min}}=100$ s, and a power-law distribution of event sizes with a slope of $\alpha=-2.5$. \textit{Right}: a strand with a half length of $\text{L}=80$ Mm, energy flux of $\text{F}=2\times10^7\ \text{erg}\ \text{cm}^{-2}\ \text{s}^{-1}$, minimum delay between events of t$_{\text{min}}=1000$ s, and a power-law distribution of event sizes with a slope of $\alpha=-1$. The \textit{top} panels indicate the volumetric heating rate, the \textit{middle} panels indicate the coronal electron temperature, and the \textit{bottom} panels indicate the coronal electron density. The gray vertical lines mark the end of the equilibration period after which the runs are averaged.}
	\label{fig:runs:timeseries}
\end{figure*}

Due to EBTEL's speed, we are able to simulate a large amount of solar time in relatively little computational time for this study. Each EBTEL model is run for $10^5$ s of solar time and $1000$ models with random realizations of impulsive heating are run for each set of parameters to provide a robust average and standard deviation. In total, $1.6\times10^9$ s of coronal loop evolution are simulated. Those EBTEL parameters that remain constant across all simulations are listed in table \ref{tab:fixed}.

The evolution of two of these models is shown in Figure \ref{fig:runs:timeseries}. For each of these models, the plasma undergoes many heating and cooling cycles in a single run. Some notable (and expected) features of these simulations include: the typically smaller, more frequent heating events in the model with the shorter minimum delay and steeper distribution of event sizes; the more consistent plasma temperature and density resulting from these more consistent heating events; the more rapid cooling in the shorter strand; and the higher plasma temperatures in the more strongly heated strand with larger heating events.

While the plasma in these models is evolving on individual magnetic strands, the observable signatures of this heating are due to the combination of many hundreds or thousands of such strands evolving within a single resolution element. In addition, because each of these strands is evolving in isolation (due to the extremely high ratio of parallel to perpendicular heat conduction along the magnetic field \citep{vandenOord1994}), the time average of the evolution of a single strand is equivalent to the average of a snapshot of many strands at different phases of their heating and cooling cycles. Because of this equivalence, we not only average all 1000 runs with each set of parameters together, we also average each run over the duration of its evolution, except for the first $10^{4}$ s that are discarded to ensure that the initial conditions of each run have no impact on the results.

\begin{figure}[htp!]
	\includegraphics[trim=0cm 2.05cm 0cm 0cm, width=\columnwidth]
{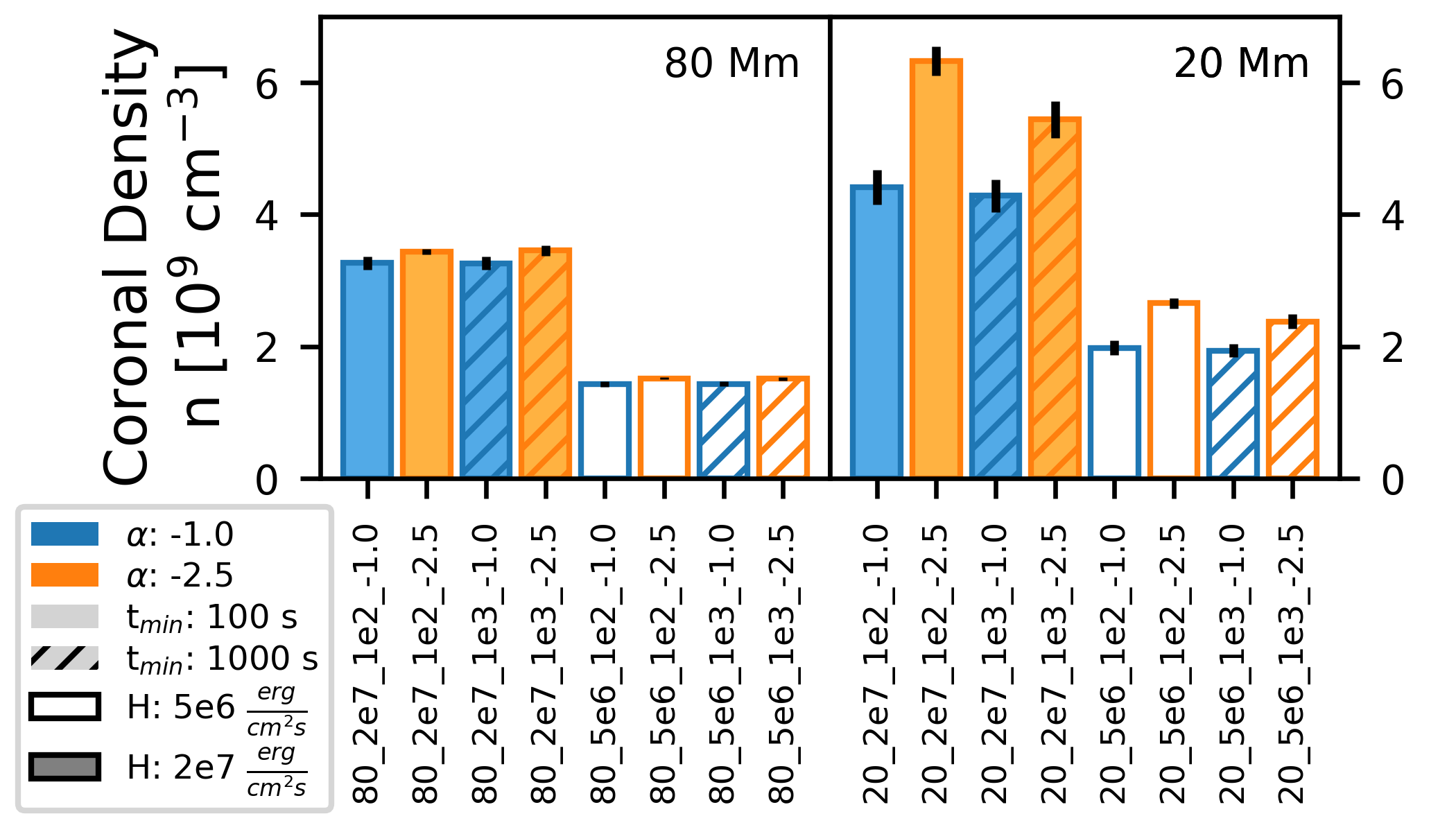}
	\vspace{0.1cm}
	\includegraphics[trim=0cm 0cm 0cm 0cm, width=\columnwidth]
{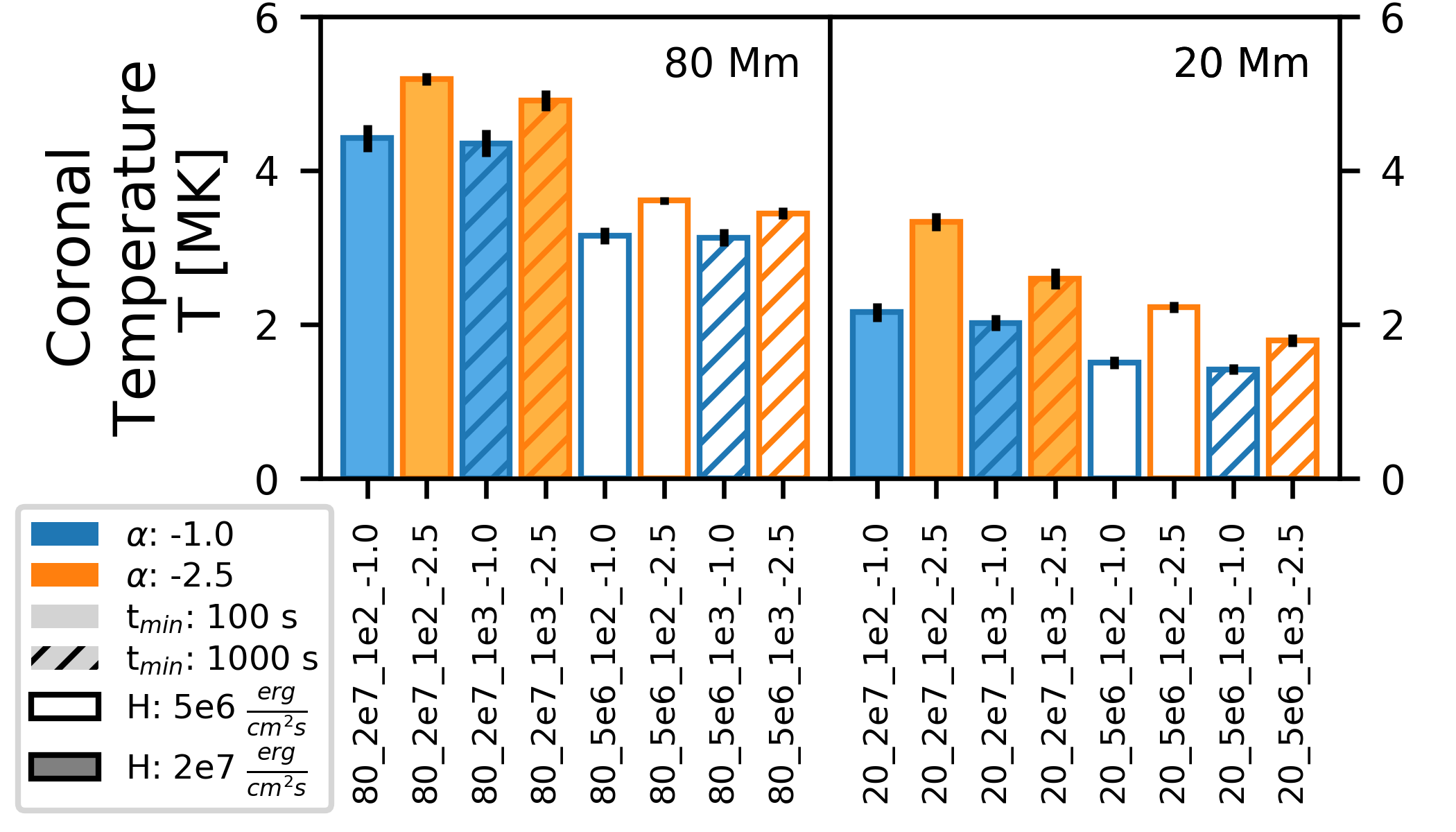}
	\caption{Average coronal plasma density (\textit{top}) and temperature (\textit{bottom}) for the 16 tested combinations of the strand parameters. Each simulation is labeled and also indicated by the combination of location (left or right panel), color (blue or orange), pattern (solid or stripped), and shading (filled or empty). The black error lines at the top of each bar indicate the standard deviation as determined by considering the time average of each of the 1000 model runs as a single sample.}
	\label{fig:runs:averages}
\end{figure}

\begin{deluxetable*}{cccccc}
  \tablewidth{0pt}
  \tablecaption{High-frequency heating models compared with loop equilibrium scaling laws}
  \tablecolumns{5}
  \tablehead{\colhead{L [Mm]} & \colhead{F [erg cm$^{-2}$ s$^{-1}$]} &
  	\colhead{$T$ [MK]} & \colhead{$\bar T$ theory [MK]} &
  	\colhead{$n$ [$10^9$ cm$^{-3}$]} &
  	\colhead{$n$ theory [$10^9$ cm$^{-3}$]}}
  \startdata
  	$80$ & $2\times10^7$ & $5.19\pm0.08$ & $5.21$ & $3.43\pm0.04$ & $3.40$ \\
  	$80$ & $5\times10^6$ & $3.61\pm0.05$ & $3.45$ & $1.52\pm0.02$ & $1.45$ \\
  	$20$ & $2\times10^7$ & $3.33\pm0.12$ & $3.45$ & $6.32\pm0.22$ & $6.30$ \\
  	$20$ & $5\times10^6$ & $2.23\pm0.07$ & $2.26$ & $2.65\pm0.08$ & $2.76$ \\
  \enddata
  \tablecomments{Temperature and density scaling of EBTEL models with $\text{t}_{\text{min}}=100$ s and $\alpha=-2.5$ compared with theoretical predictions of steady state equilibrium loops. The EBTEL modeled temperature ($T$) and density ($n$) are compared with the theoretical temperature ($\bar T$ theory) and density ($n$ theory) determined for the last three models by applying the scaling laws in reference to the first model.\label{tab:scaling}}
\end{deluxetable*}

The average density and temperature for each set of modeled parameters is given in Figure \ref{fig:runs:averages}. We can begin to understand the trends by examining the models with the highest frequency of heating events, which most closely resemble steady heating. These are the cases with the shortest minimum delay times ($\text{t}_{\text{min}}=100$ s) and steepest distributions ($\alpha=-2.5$). There are well-known scaling laws for loops with truly steady heating, one of which is \citep{Porter1995}:
\begin{equation}
\label{eqn:temperature_scaling}
\bar T\propto \text{L}^{4/7}\text{Q}^{2/7}\propto \left(LF\right)^{2/7}
\end{equation}
where $\text{Q}=\text{F}/\text{L}$ is the volumetric heating rate. The density of that same loop scales as:
\begin{equation}
\label{eqn:density_scaling}
n\propto L^{-3/7}F^{4/7}
\end{equation}
assuming a radiative loss function with power-law slope $\beta = -0.5$ \citep{Rosner1978}. Equations \ref{eqn:temperature_scaling} and \ref{eqn:density_scaling} are often presented with the apex values, $T_{a}$ and $n_{a}$, which have the same scaling but slightly different constants of proportionality. We fit the four high-frequency heating models (t$_{\text{min}}=100$ s and $\alpha=-2.5$) with linear regressions between the modeled and theoretical values to determine that the constants of proportionality in equations \ref{eqn:temperature_scaling} and \ref{eqn:density_scaling} are $0.013$ and $0.016$, respectively. We then apply these scaling laws to the same models and compare the theoretical average temperatures and densities with the averages determined from the simulations in table \ref{tab:scaling}. This shows that these EBTEL simulations with the highest frequency heating agree quite well with the equilibrium loop scaling laws. Differences can be attributed in part to differences in the radiative loss function; EBTEL uses a piecewise-continuous $\beta$ rather than a single value for all temperatures. Models with lower event frequency (longer minimum delay and shallower distributions) have lower average temperatures and densities than the corresponding higher frequency runs. At first this might seem surprising, since high energy events that occur less often produce higher peak temperatures, such as seen in Figure \ref{fig:runs:timeseries}. However, the strands cool quickly at these high temperatures and spend the majority of their time in a much cooler state, also characterized by lower density. This dominates the time averages.

\subsection{Predicting AIA intensities}
\label{sec:simulation:AIA}
\begin{figure}[t]
	\includegraphics*[trim=0.2cm 0.2cm 0cm 0.25cm, width=\columnwidth]
{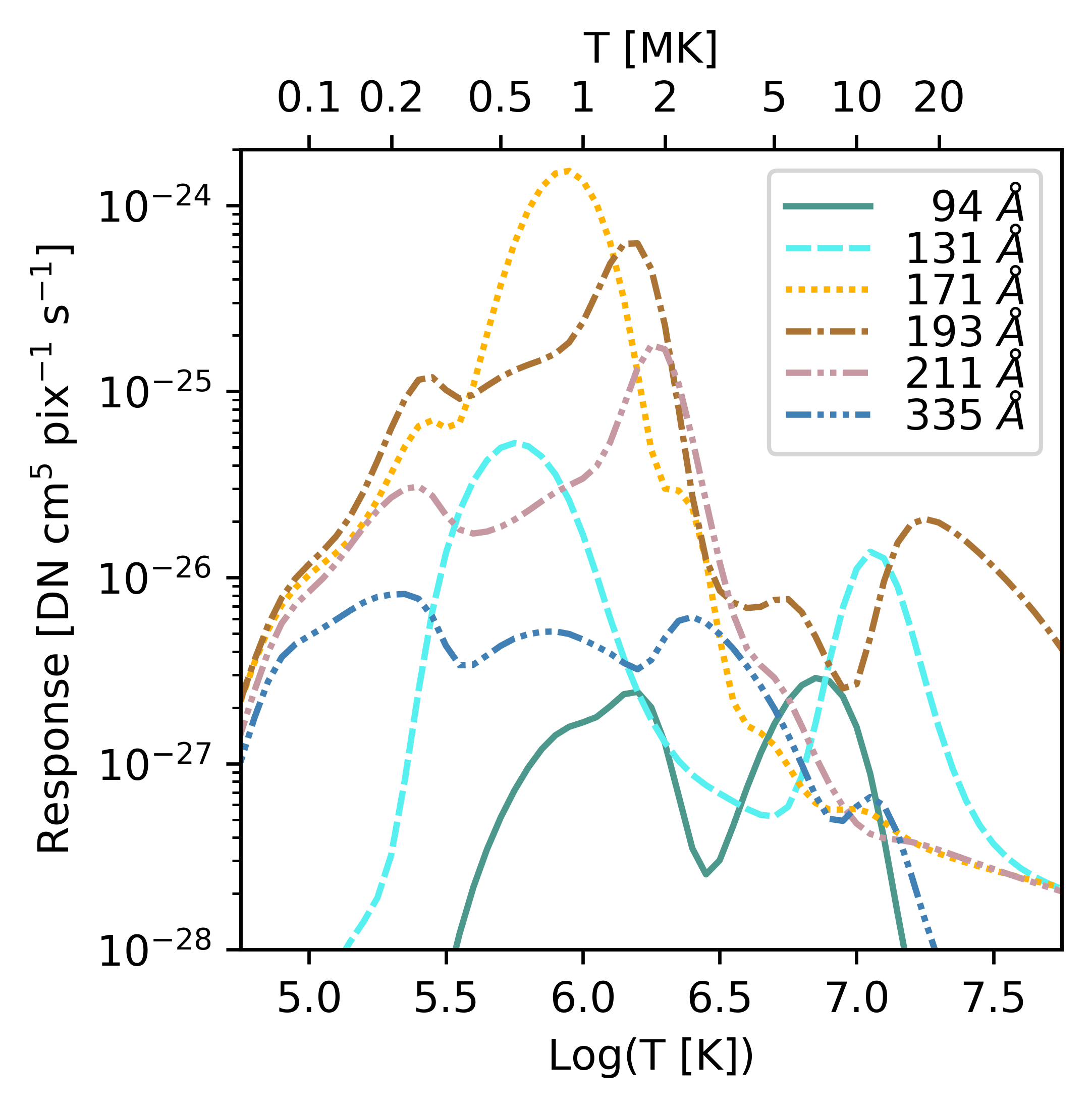}
	\caption{Temperature response functions of the six ``coronal'' AIA imaging channels. Note that the temperature response of the $171$ \AA, $193$ \AA, and $211$ \AA\ channels is concentrated near a single temperature (quasi-isothermal), while the $94$ \AA, $131$ \AA, and $335$ \AA\ channels have significant response at two or more temperatures.}
	\label{fig:aia:response}
\end{figure}

Two of the standard products of the EBTEL simulations are the time dependent DEMs of the corona and transition region. These are the plasma density squared as a function of temperature integrated through their respective portion of the modeled atmosphere. EBTEL assumes the coronal DEM at any given time is narrowly and uniformly distributed around the average coronal temperature in the strand ($\bar T$). The transition region DEM is spread between $T_{0}= 0.6 T_{a} = 0.67 \bar T$ and chromospheric temperatures and has a form determined by the energy balance between thermal conduction, radiation, and enthalpy. Using the DEMs, we can simulate the expected EUV intensity from each component of the atmosphere. We use the temperature response functions of the ``coronal'' AIA channels shown in Figure \ref{fig:aia:response} to compute the average coronal and transition region intensities of the simulated strands. These response functions are generated using the IDL routine \texttt{aia\_get\_response.pro} version 8 that utilizes version 7.1.3 of the CHIANTI atomic line database \citep{Dere1997, Landi2013}.

\begin{figure}[htp!]
	\includegraphics[trim=0cm 2.05cm 0cm 0cm, width=\columnwidth]
{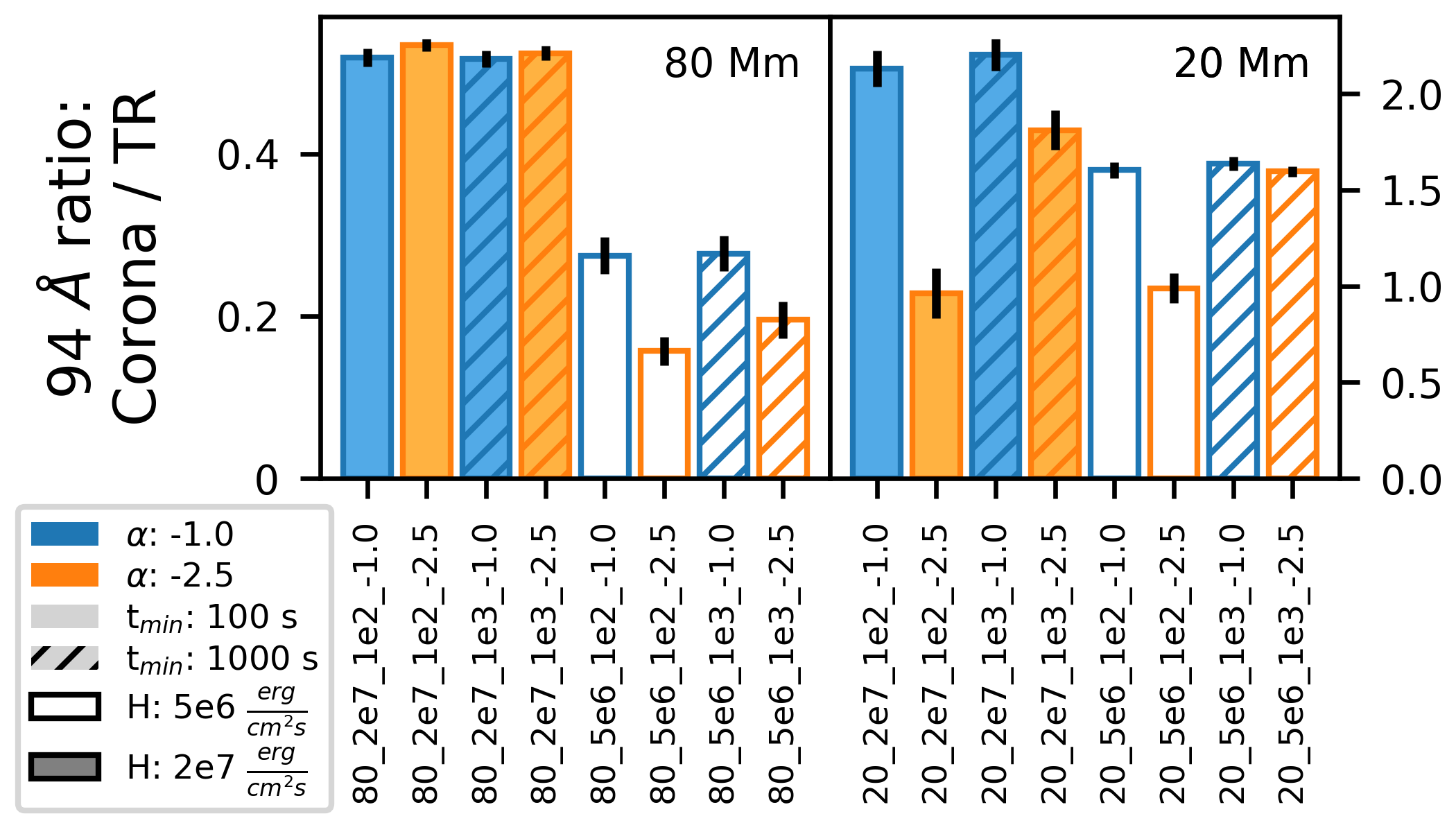}
	\vspace{0.1cm}
	\includegraphics[trim=0cm 2.14cm 0cm 0cm, width=\columnwidth]
{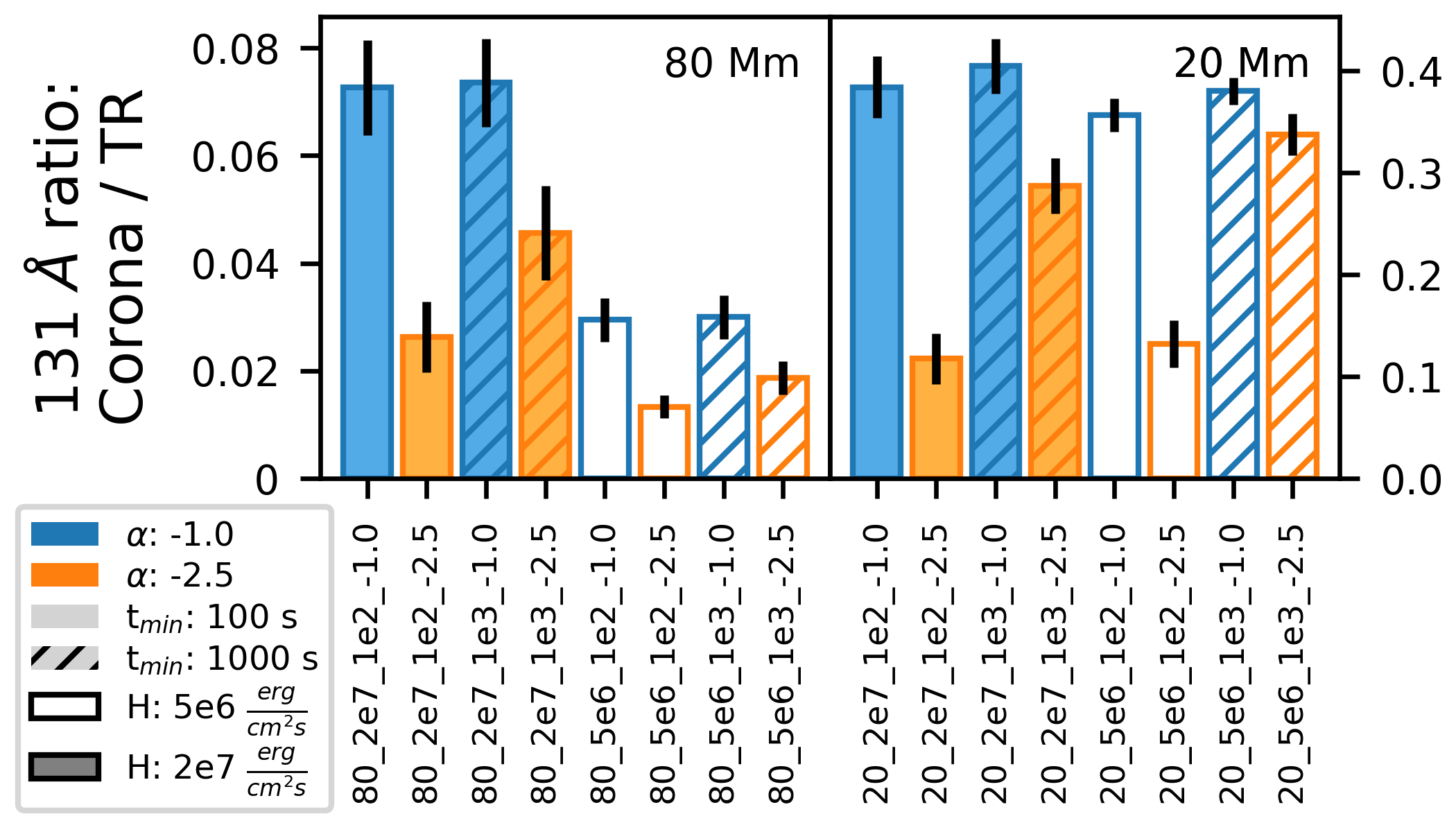}
	\vspace{0.1cm}
	\includegraphics[trim=0cm 2.14cm 0cm 0cm, width=\columnwidth]
{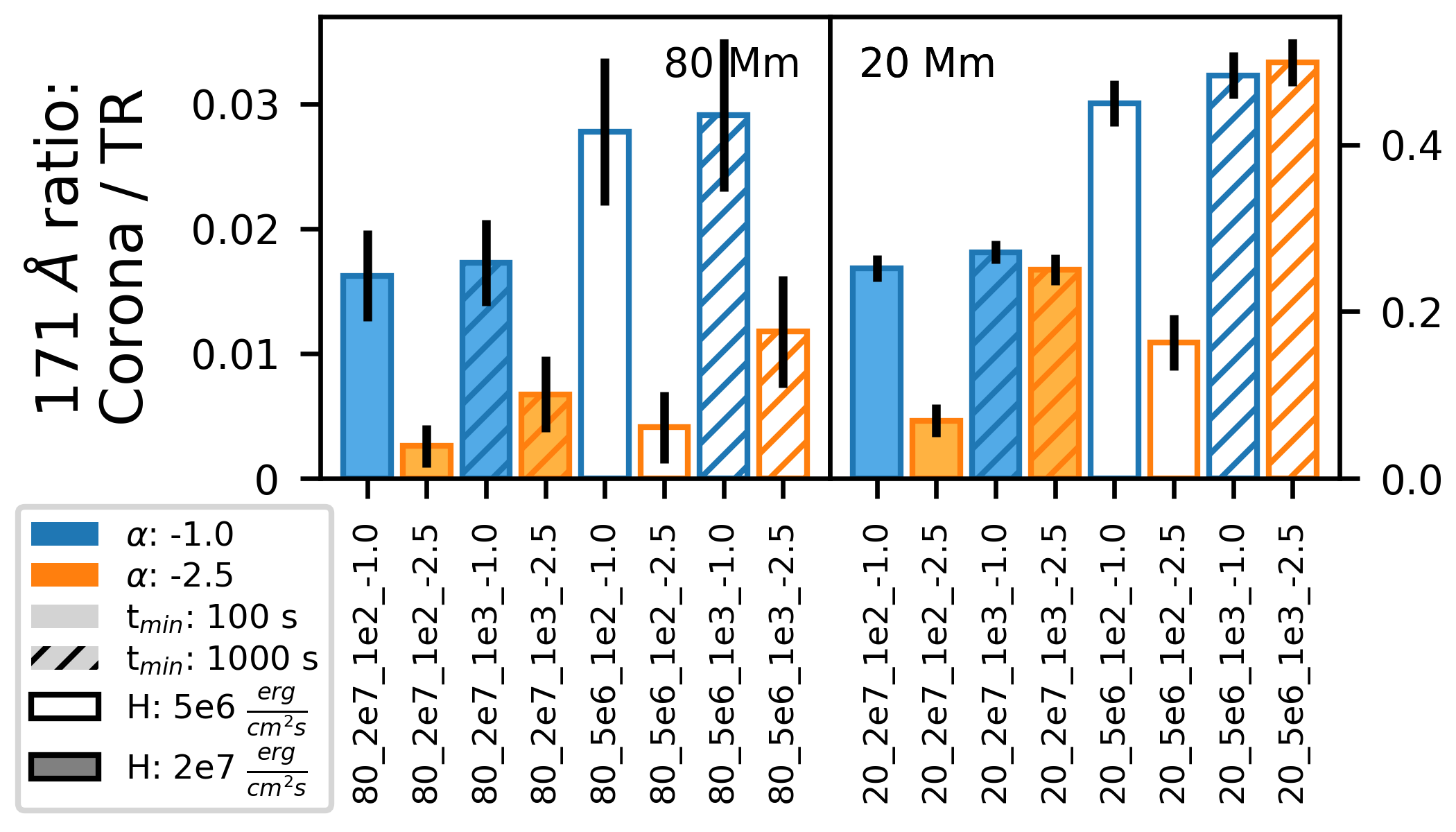}
	\vspace{0.1cm}
	\includegraphics[trim=0cm 2.14cm 0cm 0cm, width=\columnwidth]
{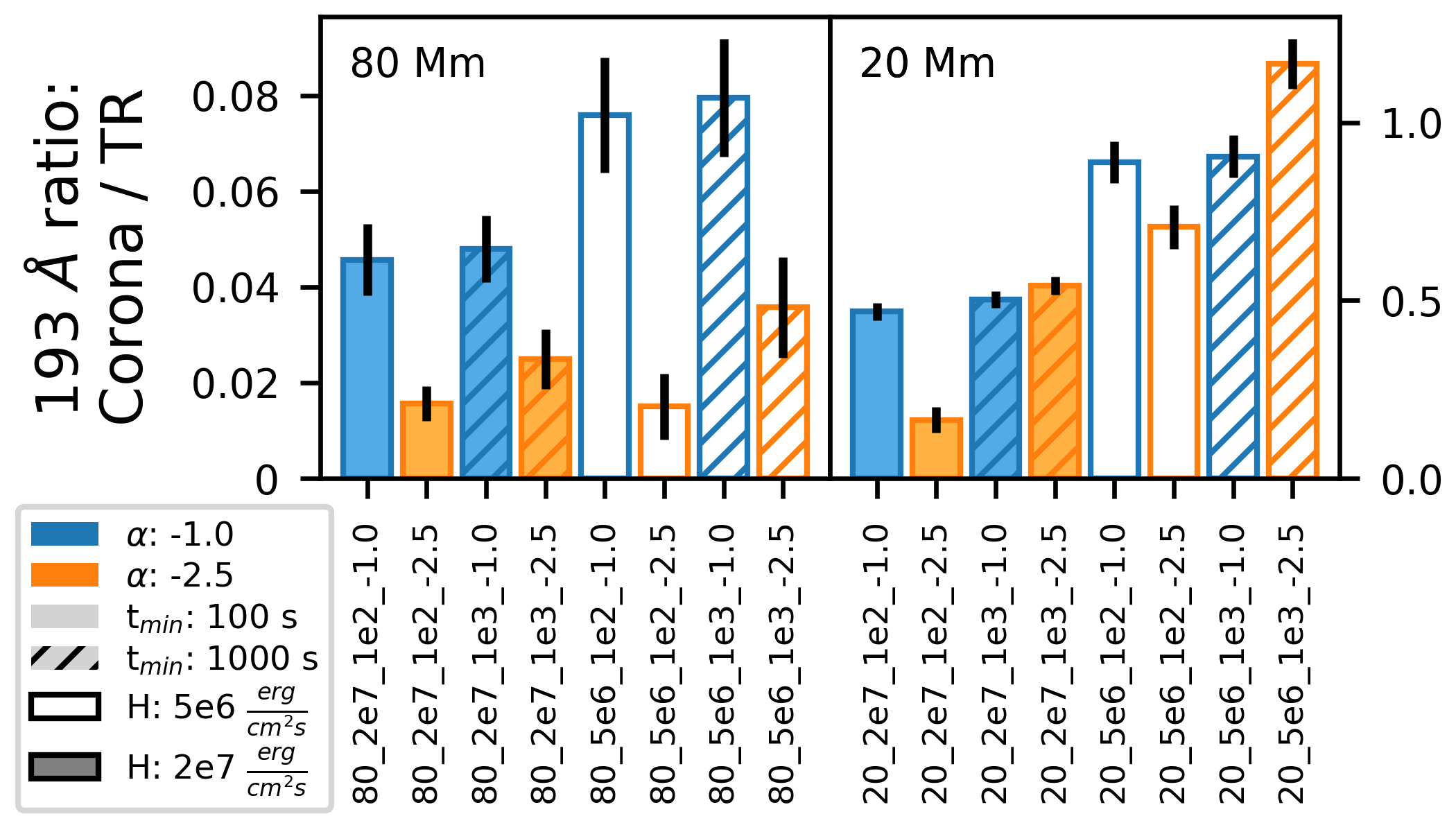}
\vspace{0.1cm}
	\includegraphics[trim=0cm 2.14cm 0cm 0cm, width=\columnwidth]
{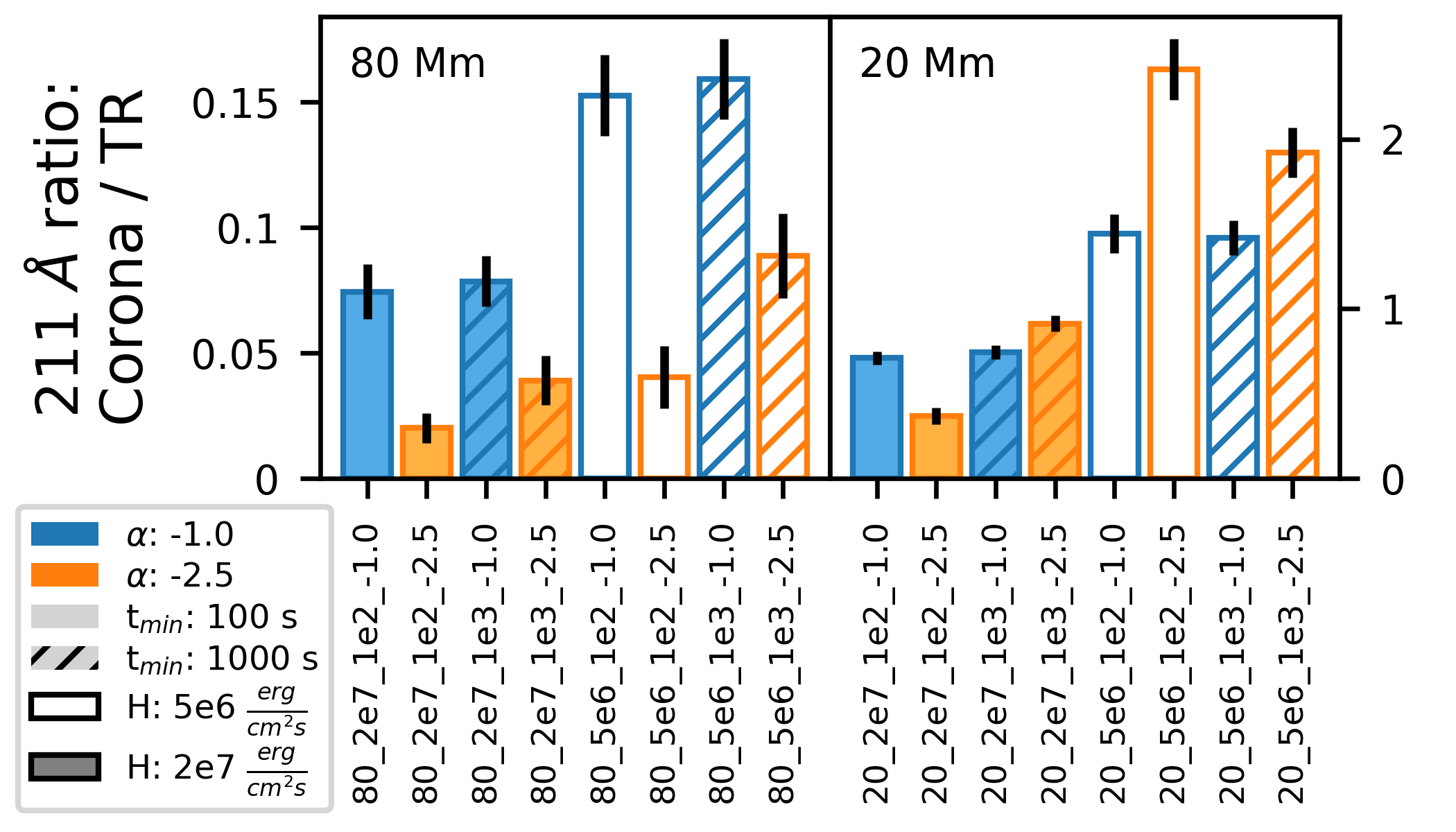}
	\vspace{0.1cm}
	\includegraphics[trim=0cm 0cm 0cm 0cm, width=\columnwidth]
{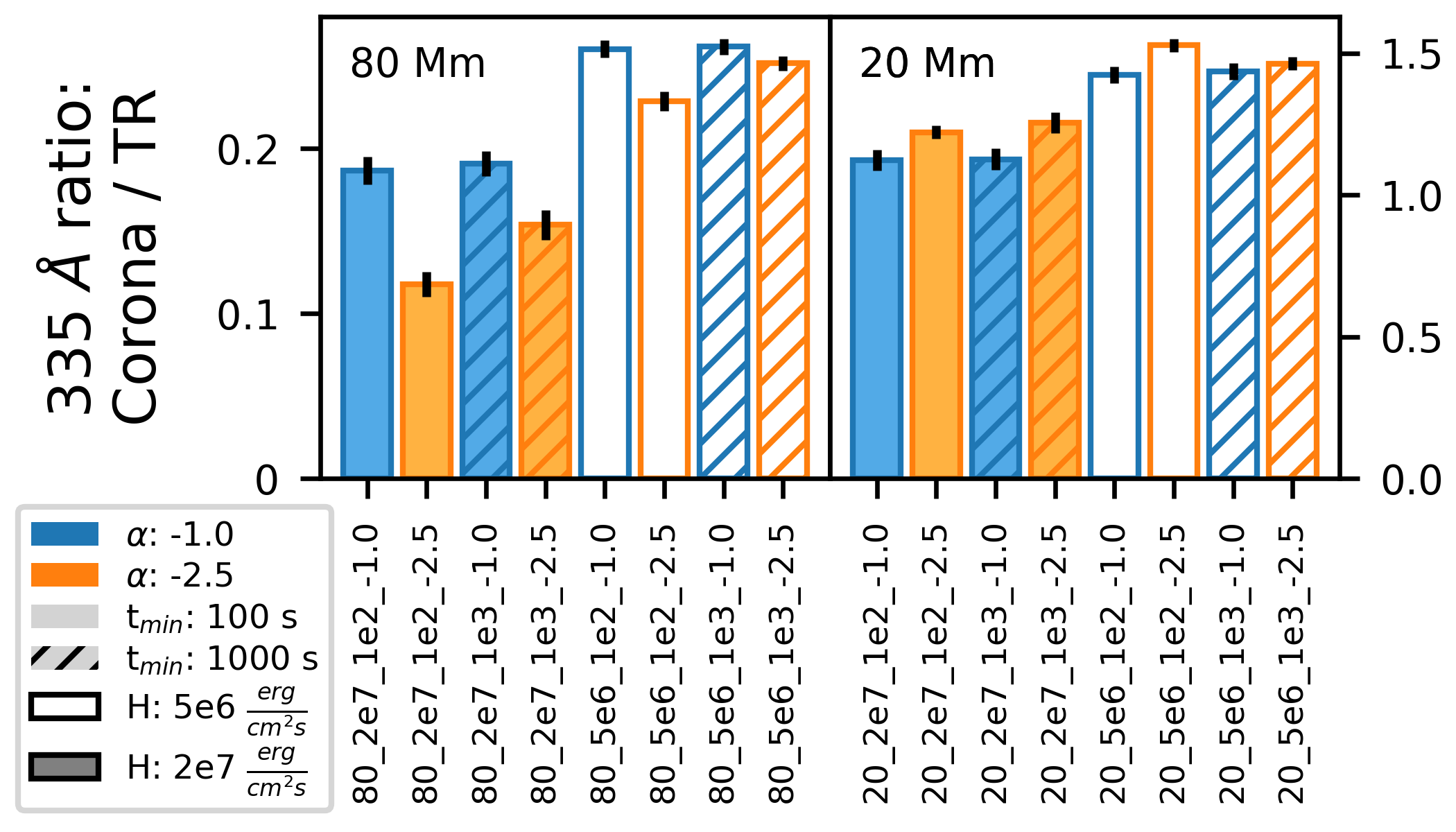}
	\caption{Ratios of coronal to transition region emission in six ``coronal'' AIA channels for the 16 tested combinations of the strand parameters. Each simulation is labeled and also indicated by the combination of location (left or right panel, with different scales), color (blue or orange), pattern (solid or stripped), and shading (filled or empty). The black error lines at the top of each bar indicate the standard deviation in the ratio as determined by considering the time average of each of the 1000 model runs as a single sample. Note that the ratios for the $20$ Mm strands are much larger than the ratios for the $80$ Mm strands.}
	\label{fig:runs:ratios}
\end{figure}

The emission from a single magnetic strand is of course different from what is observed in a pixel of a high-resolution image. The line-of-sight corresponding to that pixel passes through many different strands of differing lengths. This introduces complexity in interpreting real observations, which we return to in Section \ref{sec:observations}. For now, we take a simplified approach in order to investigate general behavior. We assume that the emission along a given line of sight can be represented by a single time-averaged strand, i.e., one of our models. A schematic drawing of the basic idea is shown in Figure 1 of \cite{Klimchuk2014}. To facilitate an approximate comparison with observations, we take the line-of-sight depth of the corona to be $40$ Mm, comparable to a typical active region scale height, and integrate the coronal DEM from the model over this length to determine the coronal intensities. We then divide this by the transition region intensity from the same model to get a corona-to-transition region intensity ratio, R$_{\text{C/TR}}$. The results are shown in Figure \ref{fig:runs:ratios} for all six ``coronal'' AIA channels and all 16 models. Note that the models with strand half lengths of $20$ Mm and $80$ Mm, corresponding to semicircular apex heights of $\approx13$ and $\approx51$ Mm, are normalized by the same $40$ Mm coronal depth. Differences in coronal brightness between models are not due to differences in depth.

The results from Figure \ref{fig:runs:ratios} yield the following general trends. In interpreting these trends, it is important to keep two things in mind. First, at any given time during the evolution of a strand, the transition region temperature extends to more than half of the apex (maximum) temperature in the strand (equation \ref{eqn:c3}). Second, the classification of heating frequency into high, intermediate, and low is based on the delay between successive heating events relative to the plasma cooling time.

\begin{itemize}
	\item In all cases, the ratio is much larger in the $20$ Mm strand than the $80$ Mm strand. This is due partly to the $40$ Mm coronal depth scaling described above. The coronal intensity used in the ratio is over and under represented in the short and long strands, respectively, compared to the full strand length simulated with EBTEL. There is an additional real effect. During a low to intermediate frequency heating and cooling cycle, the transition region emits in a narrow temperature band centered on $T$ for the entire time that the apex is cooling from its peak value to approximately $2T$. The corona, on the other hand, emits at this temperature only for the short time that it takes the coronal plasma to cool through the band. Strands that start their cooling from a higher peak temperature are therefore expected to have a smaller ratio of corona to transition region intensity. Longer strands tend to reach higher temperatures. With strong impulsive heating, the temperature rises to the point at which thermal conduction cooling balances the energy input. This determines the maximum apex temperature. We can estimate this temperature from $\text{Q} = \text{F}/\text{L} \propto T_a^{7/2}/\text{L}^2$, which shows that $T_a \propto (\text{FL})^{7/2}$. 
	
	\item In the $171$ \AA, $193$ \AA, $211$ \AA. and $335$ \AA\ channels, R$_{\text{C/TR}}$ is smaller with the larger energy flux, all else equal. This can also be explained by the argument above. Larger F implies hotter $T_a$, which means that the transition region radiates for longer. The $94$ \AA\ and $131$ \AA\ channels often display the opposite effect, which may be due to their second, high-temperature peaks. The $193$ \AA\ channel also has a second, high-temperature peak, but its reduced sensitivity compared to the primary peak and its very high temperature mean that it has a negligible influence on the channel response in these modeled loops and solar observations outside of flares.

	\item In general, the channels with higher temperature responses ($94$ \AA, $211$ \AA, and $335$ \AA)  have larger ratios than the channels with cooler temperature responses ($131$ \AA, $171$ \AA, and $193$ \AA). A variation of the above argument applies here. The maximum apex temperature of a strand is of course the same, regardless of the observing channel. A given $T$ that begins in the transition region at the start of cooling switches to being in the corona when the apex cools to approximately $2T$. This happens sooner for larger $T$, so the transition region emission turns off more quickly in hotter channels, and R$_{\text{C/TR}}$ is larger. Real channels are of course sensitive to a broad range of temperatures, but the basic concept applies.

	\item For cases with $\alpha = -1$, $t_{\text{min}}$ has almost no effect. This is because the energy input is dominated by larger heating events with longer delay times.

	\item For cases with $\alpha = -2.5$, $t_{\text{min}}$ has a large effect, particularly for the $20$ Mm strands. This is due to the cooling time of a $20$ Mm strand being of order $1000$ s, and therefore these small-event-weighted distributions are heated in either a high- or low-frequency regime depending on the choice of minimum cutoff. The effect for the $80$ Mm strands is less pronounced since even $1000$ s is less than the cooling time.

	\item In the $80$ Mm strands, the ratios are largest in the low-frequency heating scenarios (with the exception of the $94$ \AA\ channel in strands experiencing high energy flux). This is consistent with the findings from \cite{Patsourakos2008} that found impulsive (nonstatic equilibrium) heating produced larger corona to footpoint ratios in TRACE observations.

\item The arguments above do not apply to models with high-frequency heating, since they experience minimal cooling. Plasma that begins in the corona stays in the corona, and plasma that begins in the transition region stays in the transition region.  R$_{\text{C/TR}}$ still has a strong temperature sensitivity, but for a different reason. Higher temperature channels are better ``tuned'' to the corona than to the transition region, so the ratio is larger. A good example is the scenario with $\text{L}=20$ Mm, average energy flux of $\text{F}=5\times10^6\ \text{erg}\ \text{cm}^{-2}\ \text{s}^{-1}$, minimum delay between events of t$_{\text{min}}=100$ s, and $\alpha = -2.5$. The time-series for one of these models is shown in the left panel of Figure \ref{fig:runs:timeseries} which illustrates that the temperature is tightly constrained around the average of $2.2$ MK, just above the peak of the $211$ \AA\ channel. This set of parameters yields the largest R$_{\text{C/TR}}$ in the $211$ \AA\ and $335$ \AA\ (which also has significant sensitivity at these temperatures) channels while yielding the lowest R$_{\text{C/TR}}$ for models with the same energy flux in the other channels. 
\end{itemize}

Overall, Figure \ref{fig:runs:ratios} clearly demonstrates that EBTEL models of the solar atmosphere indicate both that the transition region contributes significantly to the intensity of AIA observations and that this contribution has strong dependence on the details of the underlying coronal heating. In the channels with strong response to the lowest temperatures, particularly $131$ \AA\ and $171$ \AA, this analysis suggests that the majority of observed emission could be due to plasma more accurately attributed to the transition region than the corona, for a wide range of loop lengths. This is also true of the hotter channels in the long loops. Furthermore, in every channel except $335$ \AA, R$_{\text{C/TR}}$ is different by more than a factor of 2 for certain combinations of minimum delay and event distribution power law for a given loop length and energy flux. While these results are difficult to apply directly to the interpretation of observational data, as explained in Section \ref{sec:observations}, they highlight the importance of considering contributions from the transition region when using observations to characterize coronal heating.

Before proceeding to consider observations, we note that \cite{Patsourakos2008} used EBTEL simulations to investigate the coronal and transition region emission as observed in the $171$ \AA\ channel of the Transition Region And Coronal Explorer \citep[TRACE;][]{Handy1999}. Their approach differs from ours in that they treated observations near the limb, assuming that the line of sight is perpendicular to the plane of the strand, and spreading the transition region emission over $2$ Mm vertically from the solar surface. They found intensity ratios of about $1/600$ and $1/35$ for steady and low-frequency impulsive heating, respectively, in a $25$ Mm (half length) strand. These ratios correspond to $\sim0.03$ and $\sim0.6$ for our assumed observing geometry ($40$ Mm coronal path lengths), consistent with what we calculate here. \cite{Patsourakos2008} emphasized how the larger $171$ \AA\ ratios produced by impulsive heating are more in line with observations.

\section{AIA observations}
\label{sec:observations}
Since the launch of the Solar Dynamics Observatory \citep[SDO;][]{Pesnell2011} in 2010, the AIA \citep{Lemen2012} has become the default imager for studies of the solar corona. However, as demonstrated in Section \ref{sec:simulation:AIA}, a significant portion of the light observed in the AIA channels may originate in the transition region rather than the corona. In the following sections, we make simplifying assumptions about the geometry of observed active regions to distinguish the observed coronal and transition region contributions to the six ``coronal'' AIA channels.

\subsection{Observationally separating the corona and transition region}
\label{sec:observations:separating}
On the Sun, a single line of sight typically passes through the coronae of one set of strands and the transition regions of an entirely different set of strands, not the corona and transition region of the same strand, as assumed in the modeling described in Section \ref{sec:simulation}. This is only a minor concern for understanding coronal heating if the strands are similar, but that is often not the case. Instead, to compare with the modeled magnetic strands, we must investigate multiple lines of sight containing observed coronal and transition region emissions that are physically linked by the magnetic field. This is possible whenever individual loops, or collections of loops, and their associated footpoint(s) can be identified in an image.

\begin{figure*}[t]
	\gridline{
	\includegraphics[trim=0.75cm 0.0cm 1.25cm 0.5cm, width = 0.4925\textwidth, clip=true]
{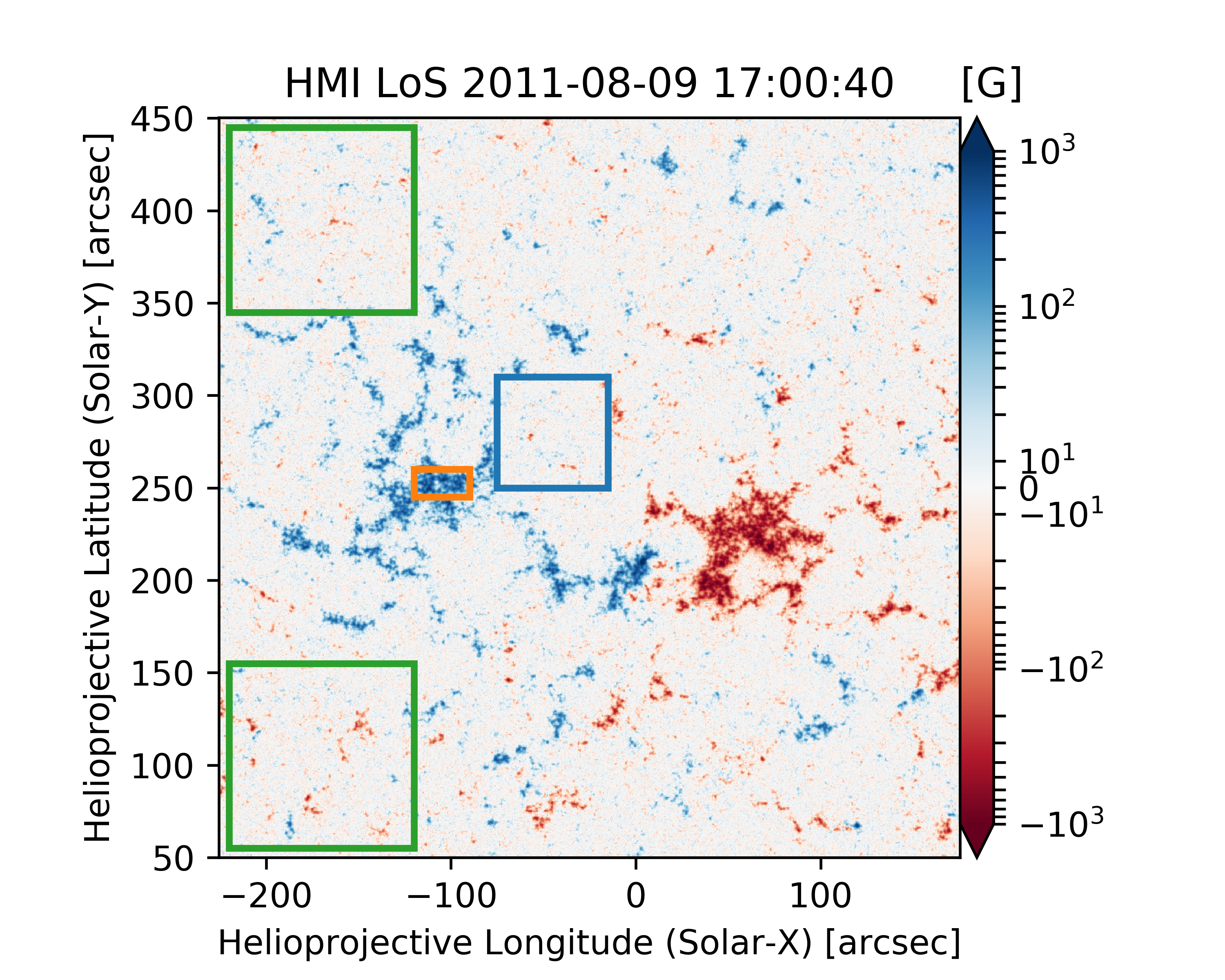}
	\includegraphics[trim=0.75cm 0.0cm 1.25cm 0.5cm, width = 0.4925\textwidth, clip=true]
{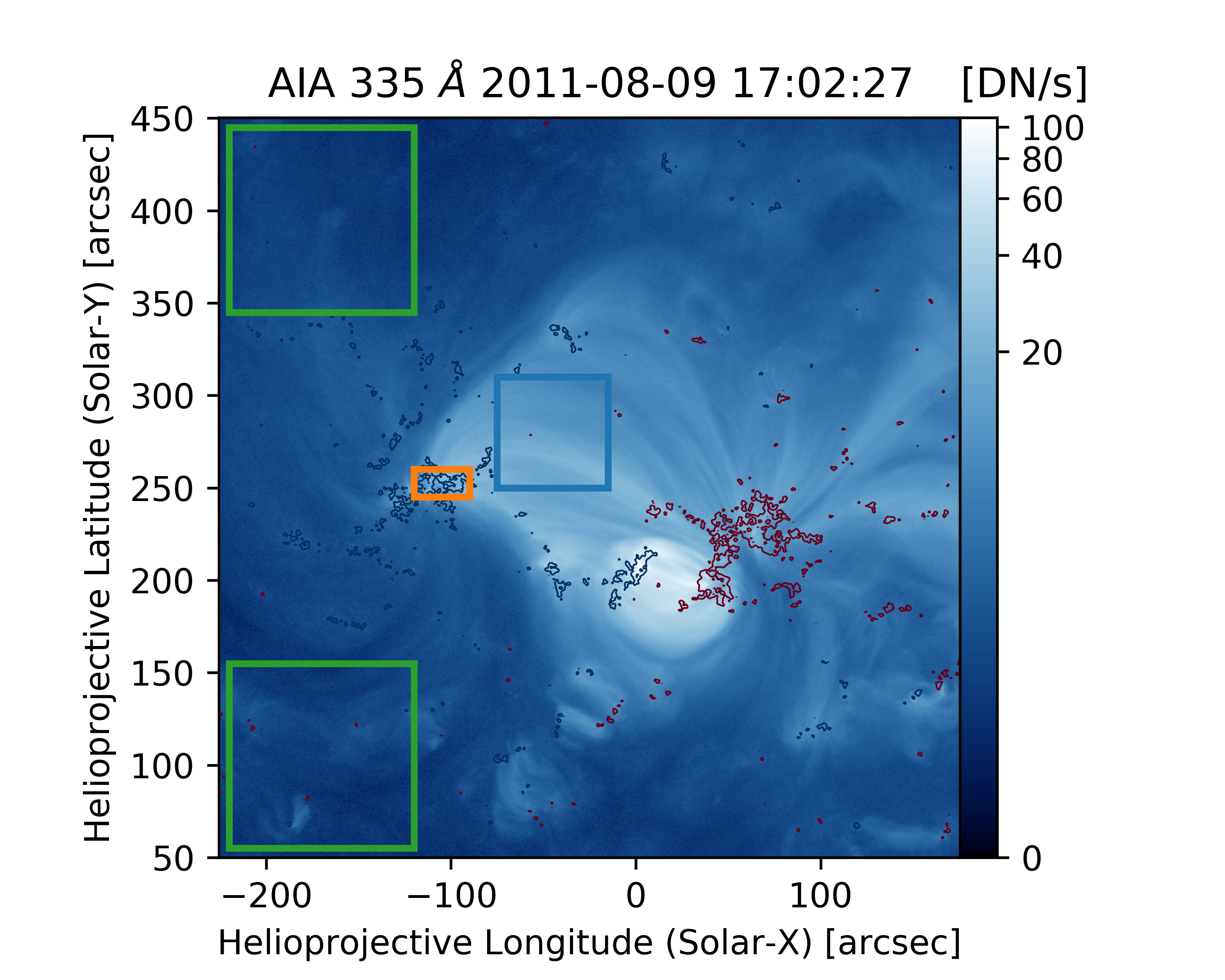}}
	\vspace{-0.4cm}
	\gridline{
	\includegraphics[trim=0.75cm 1.4cm 11.5cm 1.4cm, width = 0.011\textwidth, clip=true]
{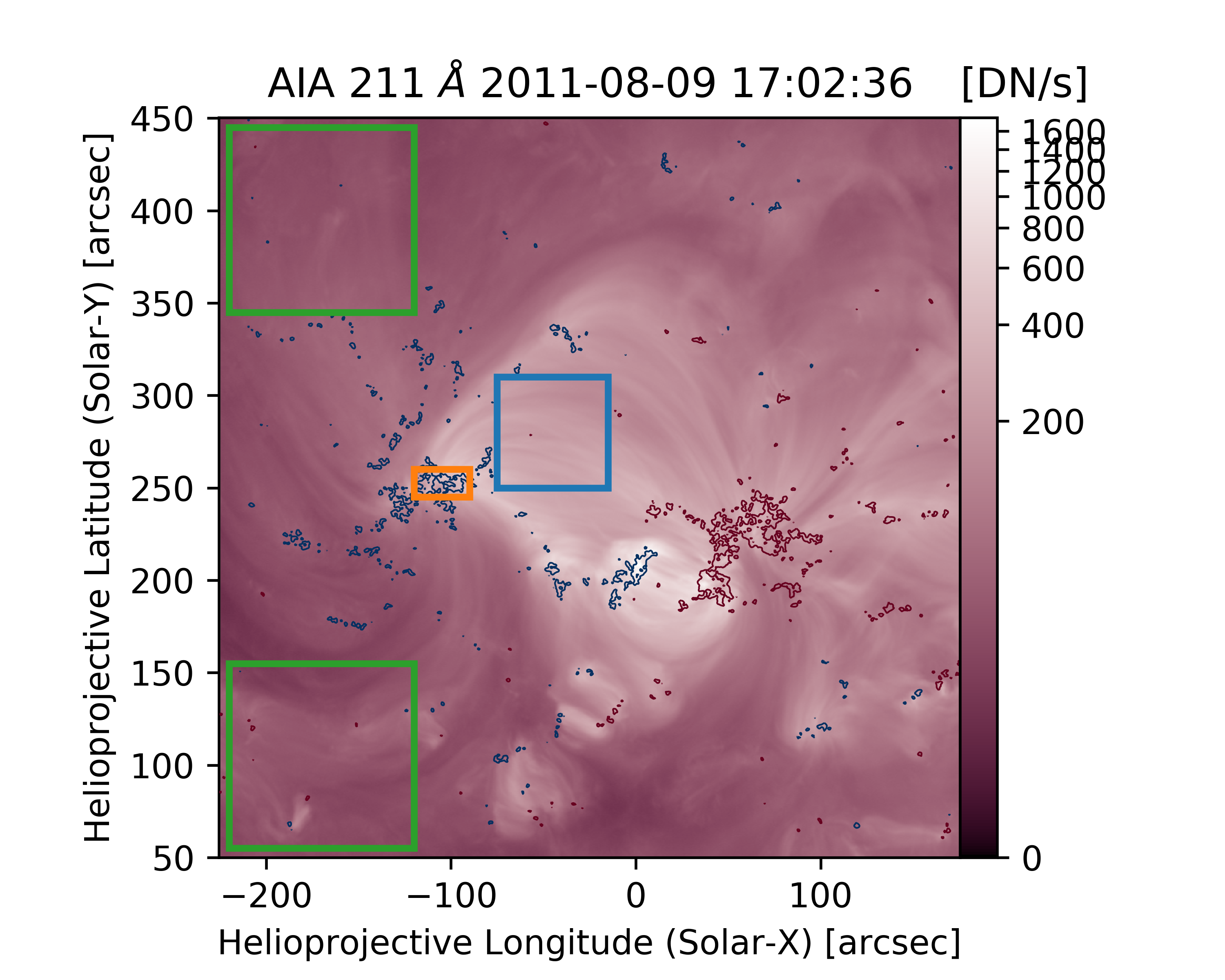}
	\includegraphics[trim=1.225cm 0.55cm 1.25cm 0.5cm, width = 0.1925\textwidth, clip=true]
{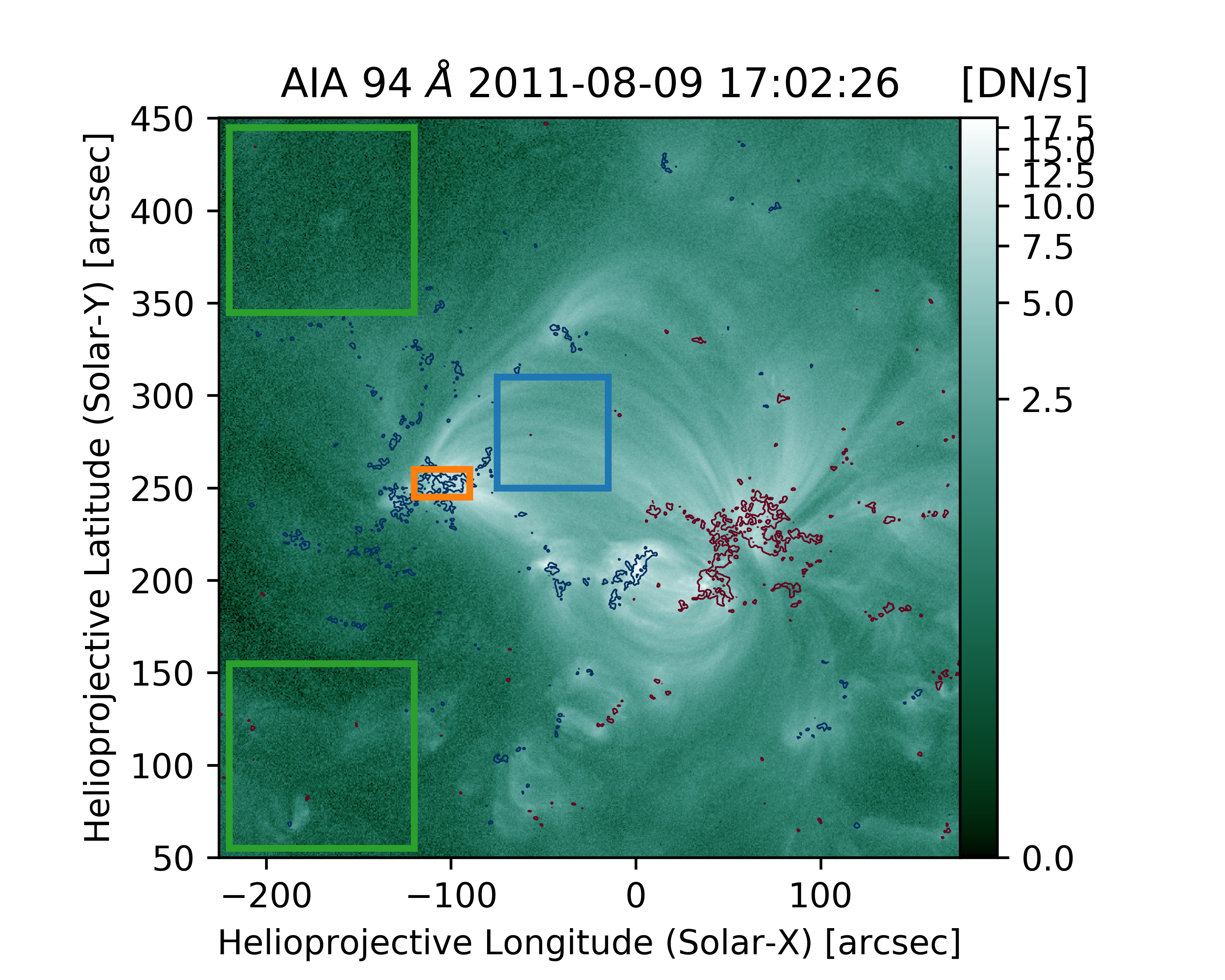}
	\includegraphics[trim=1.225cm 0.55cm 1.25cm 0.5cm, width = 0.1925\textwidth, clip=true]
{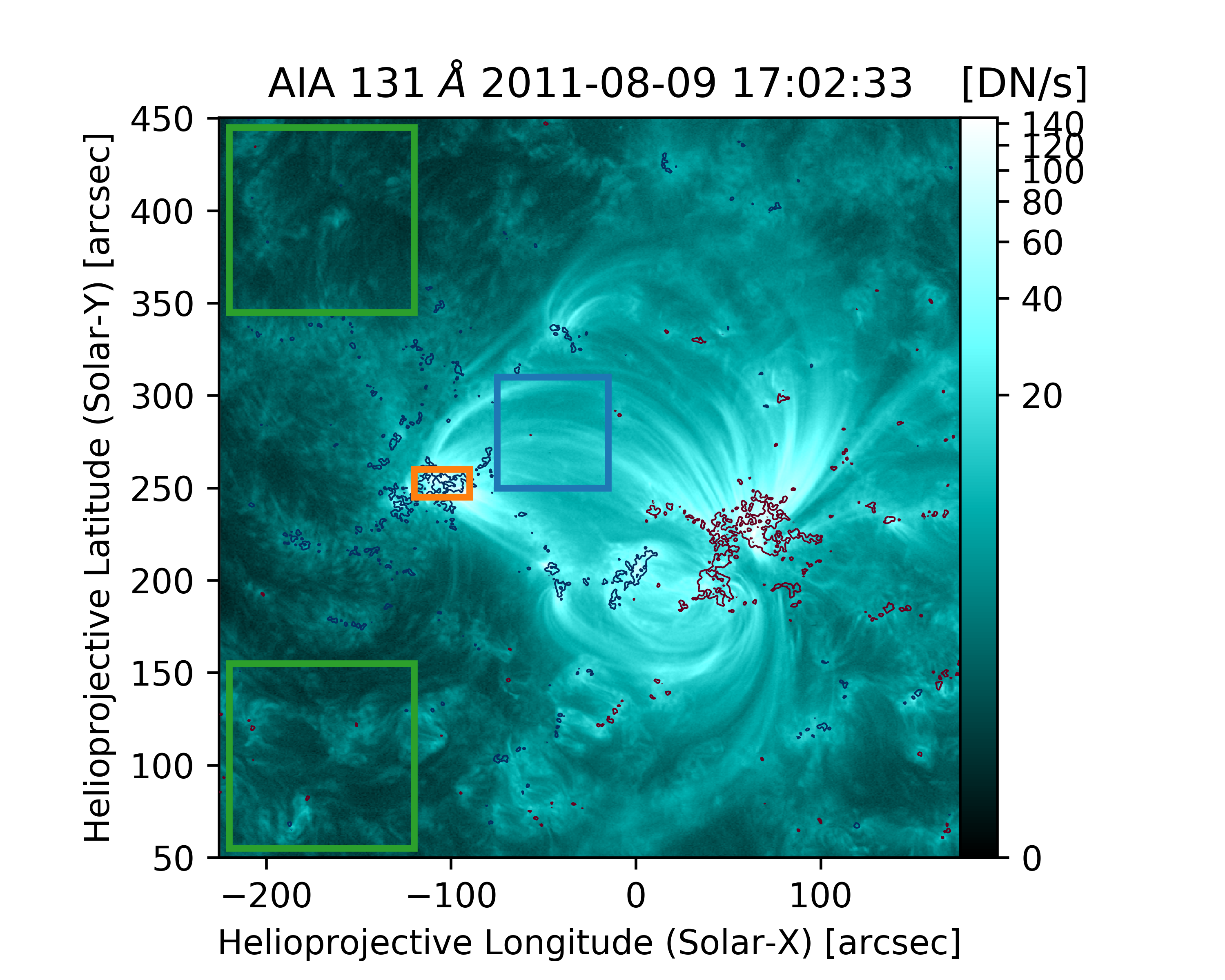}
	\includegraphics[trim=1.225cm 0.55cm 1.25cm 0.5cm, width = 0.1925\textwidth, clip=true]
{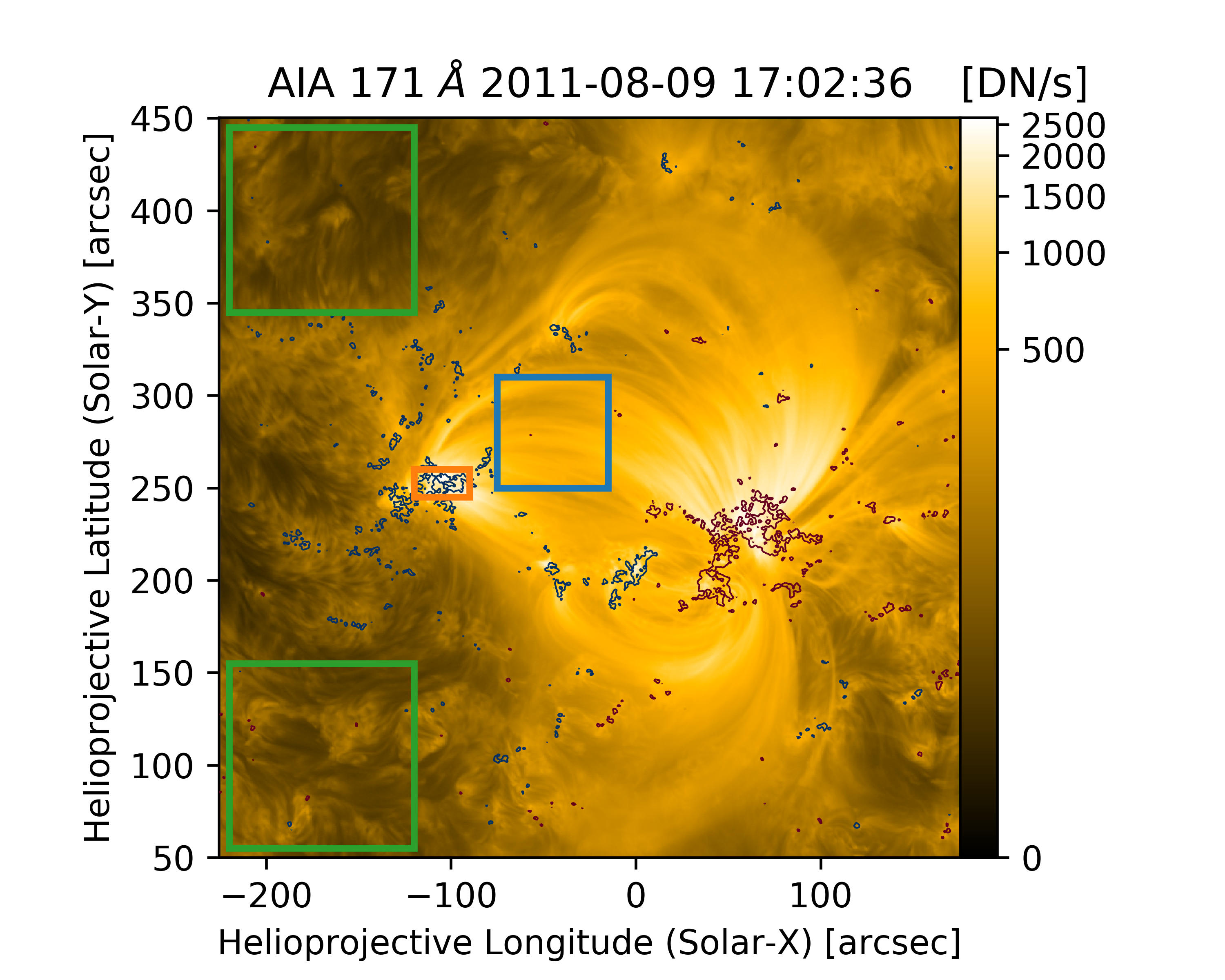}
	\includegraphics[trim=1.225cm 0.55cm 1.25cm 0.5cm, width = 0.1925\textwidth, clip=true]
{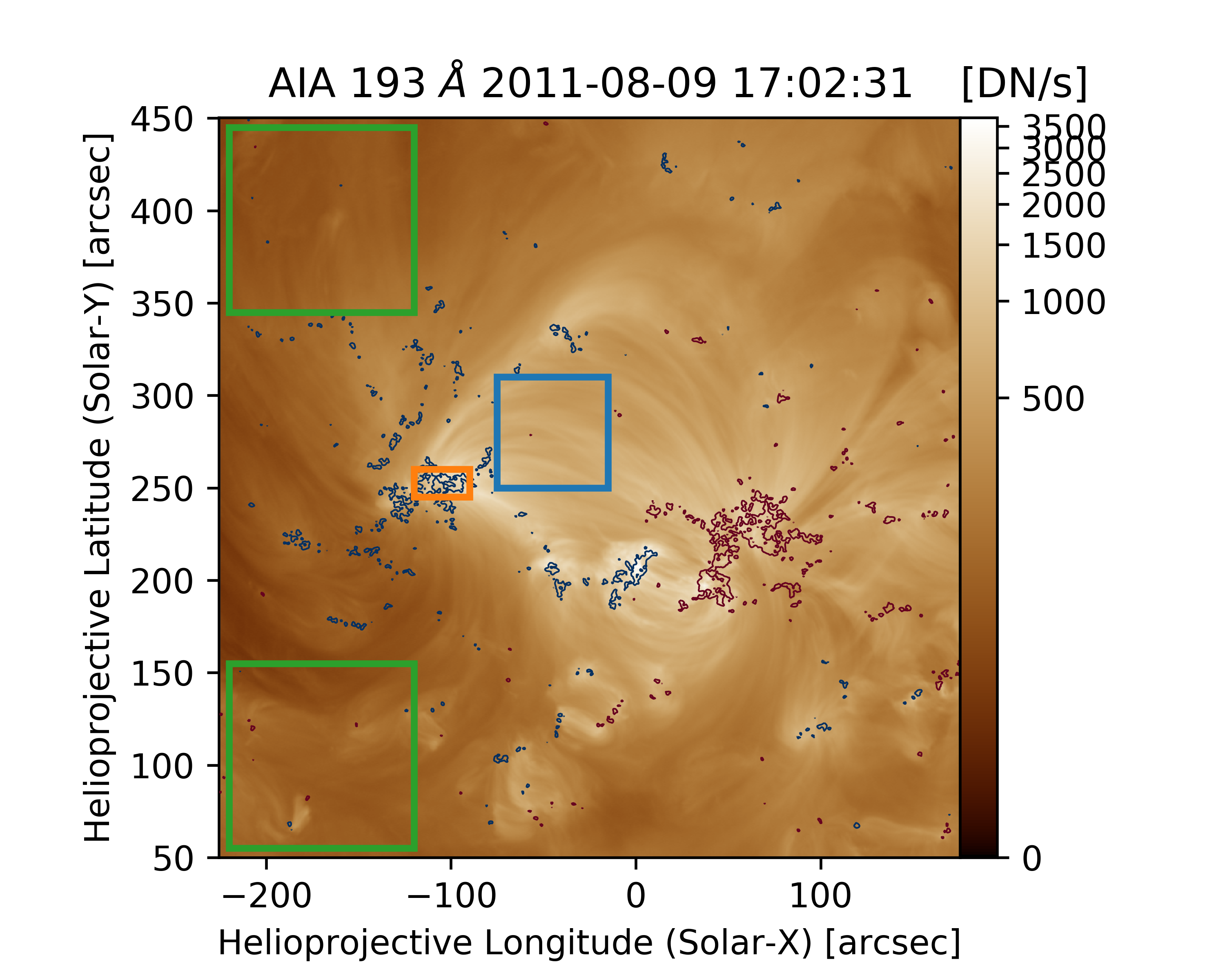}
	\includegraphics[trim=1.225cm 0.55cm 1.25cm 0.5cm, width = 0.1925\textwidth, clip=true]
{2011-08-09_211A.png}}
	\vspace{-0.6cm}
	\gridline{
	\includegraphics[trim=0.75cm 0.2cm 11.5cm 9.55cm, clip, width = 0.011\textwidth]
{2011-08-09_211A.png}  % Left axis title fudge
	\includegraphics[trim=2.25cm 0.2cm 2.5cm 9.55cm, clip, width = 0.1925\textwidth]
{2011-08-09_211A.png}  % Bottom axis title
	\includegraphics[trim=2.25cm 0.2cm 2.5cm 9.55cm, clip, width = 0.1925\textwidth]
{2011-08-09_211A.png}  % Bottom axis title
	\includegraphics[trim=2.25cm 0.2cm 2.5cm 9.55cm, clip, width = 0.1925\textwidth]
{2011-08-09_211A.png}  % Bottom axis title
	\includegraphics[trim=2.25cm 0.2cm 2.5cm 9.55cm, clip, width = 0.1925\textwidth]
{2011-08-09_211A.png}  % Bottom axis title
	\includegraphics[trim=2.25cm 0.2cm 2.5cm 9.55cm, clip, width = 0.1925\textwidth]
{2011-08-09_211A.png}}  % Bottom axis title
	\caption{The HMI line-of-sight magnetogram and the six ``coronal'' AIA channel observations of active region NOAA 11268. Each of the AIA images are five minute averages of full cadence data. The green squares indicate the regions designated as quiet sun, the blue square indicates the loop tops in the corona, and the orange rectangle indicates the footpoints and transition region of these same loops. The blue and red contours in the AIA images indicate the extent of the $\pm200$ G photospheric line-of-sight magnetic field.}
	\label{fig:aia:region}
\end{figure*}

An example for active region NOAA 11268 is shown in Figure \ref{fig:aia:region}. We select this region because of its widely separated bipole magnetic field structure with easily identifiable loops that clearly terminate in a compact concentration of strong photospheric magnetic fields. In addition, the loop top region we identify as a sample of the corona (blue box) has very weak photospheric magnetic fields along the line of sight, suggesting that there will be very little contribution from transition region plasma associated with other structures. The smaller orange box identifies the transition region footpoints that we associate with these loops. We analyze the average over five minutes of full cadence (12 s) data in order to minimize the impact of any particularly short-term variability within the region. While this average may incorporate multiple complete heating cycles (e.g. if $\text{t}_{\text{min}} = 100\text{s}$) we expect no information loss from this procedure due to the inherent averaging in the observations caused by the many overlapping and out of phase strands along a line of sight. This 5 minute averaging is consistent with the procedure from \cite{Warren2012} discussed in Section \ref{sec:observations:warren2012}.

The average intensities within the boxed regions are used to determine the characteristic coronal and transition region intensities of the prominent loops in this region. Because the photospheric magnetic field within the blue box resembles that within the quiet-Sun, we subtract the average intensity of the quiet Sun (identified by the green boxes in the upper and lower left corners) from the intensity of the corona (blue box). This has very little impact on the analysis because the quiet sun intensity is small compared to the loop intensity in these channels. We make two different assumptions about the source of the intensity in the orange box that we call the transition region. First, we assume that all of the emission comes from the transition region. Second, we acknowledge that some of the intensity is due to the overlying corona, and assume that the coronal component is identical to that in the blue box. This is likely an overestimate because we expect coronal emission to diminish from the polarity inversion line outward, both horizontally and vertically, because the heating rate varies directly with the magnetic field strength. Shorter strands tend to be brighter --- due to their increased density, as seen in Figure \ref{fig:runs:averages} --- and the line of sight intersects more short strands in the blue box than in the orange box. See Figure 1 in \cite{Klimchuk2014}.

\begin{figure}[t]
	\includegraphics*[trim= 0.25cm 0.1cm 0.8cm 0.25cm, clip, width=\columnwidth]
{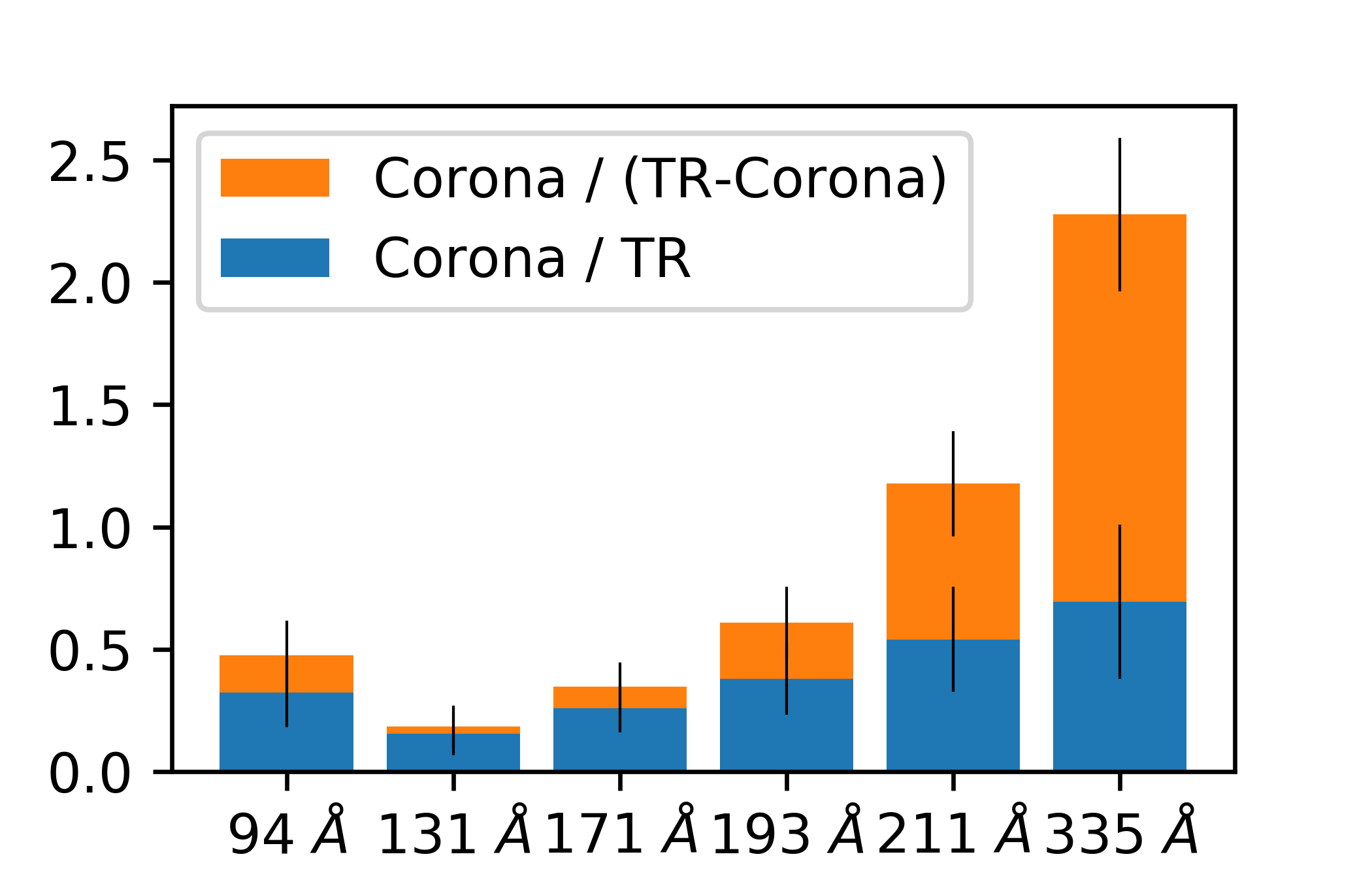}
	\caption{Observed R$_{\text{C/TR}}$ in the six ``coronal'' AIA channels for active region NOAA 11268. The blue bars represent a scenario where there is no overlying corona in the transition region (orange box) while the orange bars assume that the identified coronal intensity (blue box) is also present in the transition region. The true ratios likely fall between these two representations.}
	\label{fig:aia:region:ratios}
\end{figure}

The observed R$_{\text{C/TR}}$ using both assumptions about the contribution of the corona to the orange box is plotted for each channel in Figure \ref{fig:aia:region:ratios}. In all cases, the blue bars indicate that the transition region is brighter than the corona. When we take into account that there will be some contribution from the corona in the box identified as the transition region, the ratio increases, and significantly in the case of the 211 \AA\ and 335 \AA\ channels. This is not surprising since, in this small active region, we might expect these two relatively hotter channels to be the brightest in the corona, as can be seen in the images. In reality, the true ratios likely fall somewhere between the blue and orange bars in Figure \ref{fig:aia:region:ratios}.

While we do not anticipate any single EBTEL model will agree with the ratios observed in this active region, because it contains contributions from a large number of magnetic strands of differing length and, presumably, heating properties, it is encouraging to see the same general trends as those identified in the models. Regardless of the foreground coronal subtraction, the transition region is brighter than the corona in the $131$ \AA\ and $171$ \AA\ channels that are sensitive to lower temperature plasma and the corona is relatively brighter in the $211$ \AA\ and $335$ \AA\ channels that are sensitive to hotter plasma. The $193$ \AA\ channel samples intermediate temperatures and exhibits an intermediate ratio. The fact that the $94$ \AA\ ratio closely resembles that of $171$ \AA\ and $193$ \AA\ suggests that its emission is dominated by the low temperature peak in its temperature response function (Figure \ref{fig:aia:response}) and therefore that there is less plasma near $\sim5$ MK than near $\sim1$ MK.

\subsubsection{The impact of loop geometry}
\label{sec:observations:geometry}
In addition to the single strand/multistrand difference, there are geometrical effects that impact the comparison of the modeled and observed intensity ratios. The observed loops, or at least their envelope, appear to be considerably more compact (particularly in latitude) at their footpoints than at their apexes. Hence, the orange transition region box is smaller than the blue coronal box. The intensities that are used in the ratios are the spatial averages over the boxes. The coronal value is smaller than would be the case if all the emission were confined to a smaller area, i.e., an expanding versus nonexpanding loop. Since the models do not account for this effect, the modeled corona-to-transition region intensity ratios would need to be decreased for a more direct comparison with the observed ratios. Another geometric difference is that the models assume a coronal path length of $40$ Mm, whereas the line-of-sight depth of loops within the blue box could be larger or smaller. Finally, the coronal values from the models are the spatial averages along a strand, whereas the observed coronal intensities are near the loop apexes. Gravitational stratification would suggest that the modeled ratios should be decreased somewhat for a more direct comparison with the observations, particularly for the 80 Mm loops.

Since the modeled ratios are, if anything, too small compared to the observations, these corrections would make the discrepancy worse. However, we stress that the modeled ratios are highly idealized. The point of the present study is not to reproduce the observations as closely as possible, but rather to (1) demonstrate that the transition region makes an important contribution to intensities observed in AIA ``coronal'' channels and (2) demonstrate that R$_{\text{C/TR}}$ is sensitive to the details of the heating and therefore has diagnostic potential. In future work, we will construct more realistic models along the lines of those in e.g., \cite{Warren2006, Warren2007, Lundquist2008a, Lundquist2008b, Bradshaw2016, Nita2018, Barnes2019}.

\subsection{Analyzing the \cite{Warren2012} active regions}
\label{sec:observations:warren2012}
\begin{figure*}[t]
	\gridline{
	\includegraphics[trim=0.75cm 1.4cm 11.5cm 1.4cm, clip, width = 0.0125\textwidth]
{2011-08-09_211A.png}  % Left axis title
	\includegraphics[trim=1.225cm 0.55cm 1.25cm 0.5cm, clip, width = 0.24\textwidth]
{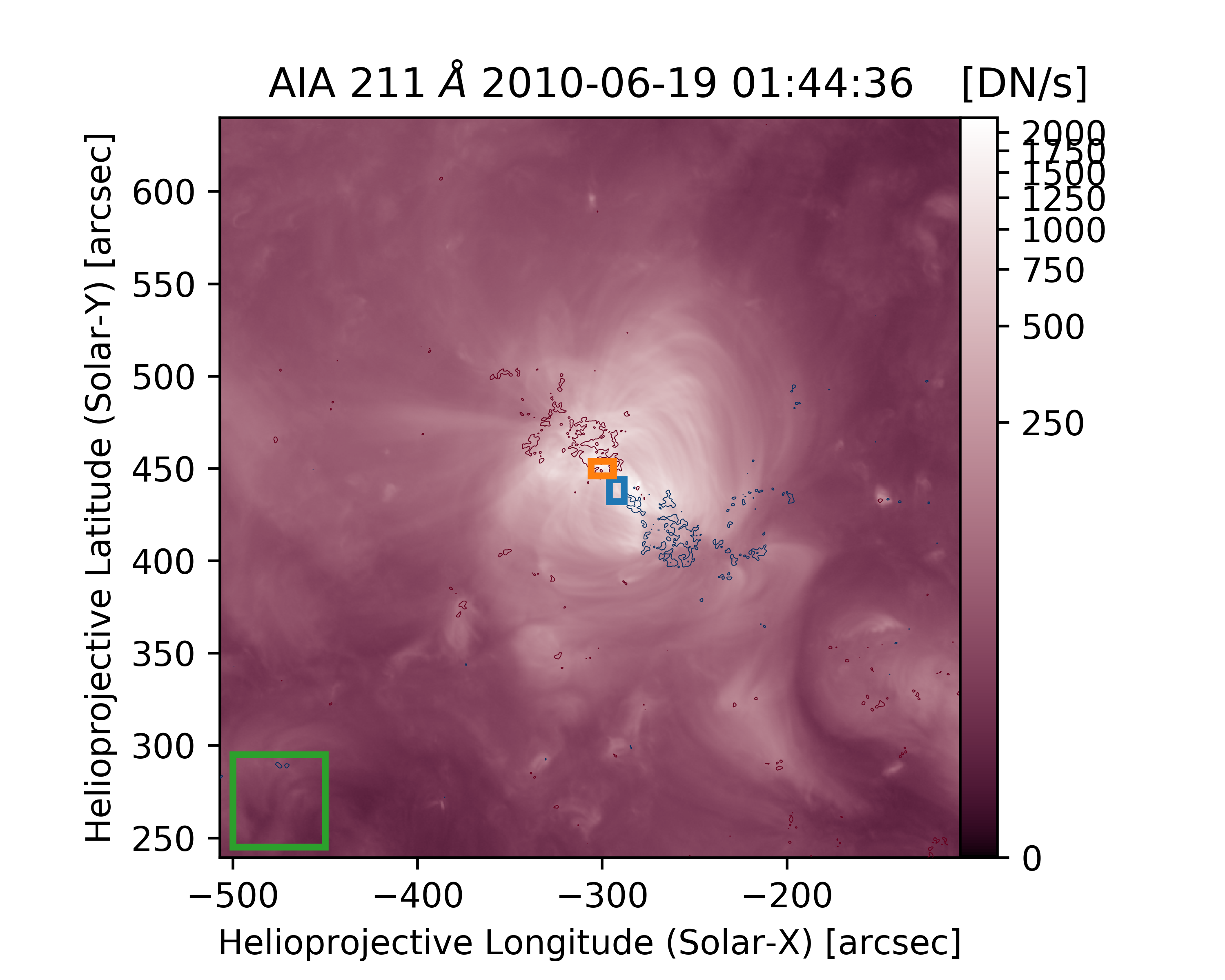}
	\includegraphics[trim=1.225cm 0.55cm 1.25cm 0.5cm, clip, width = 0.24\textwidth]
{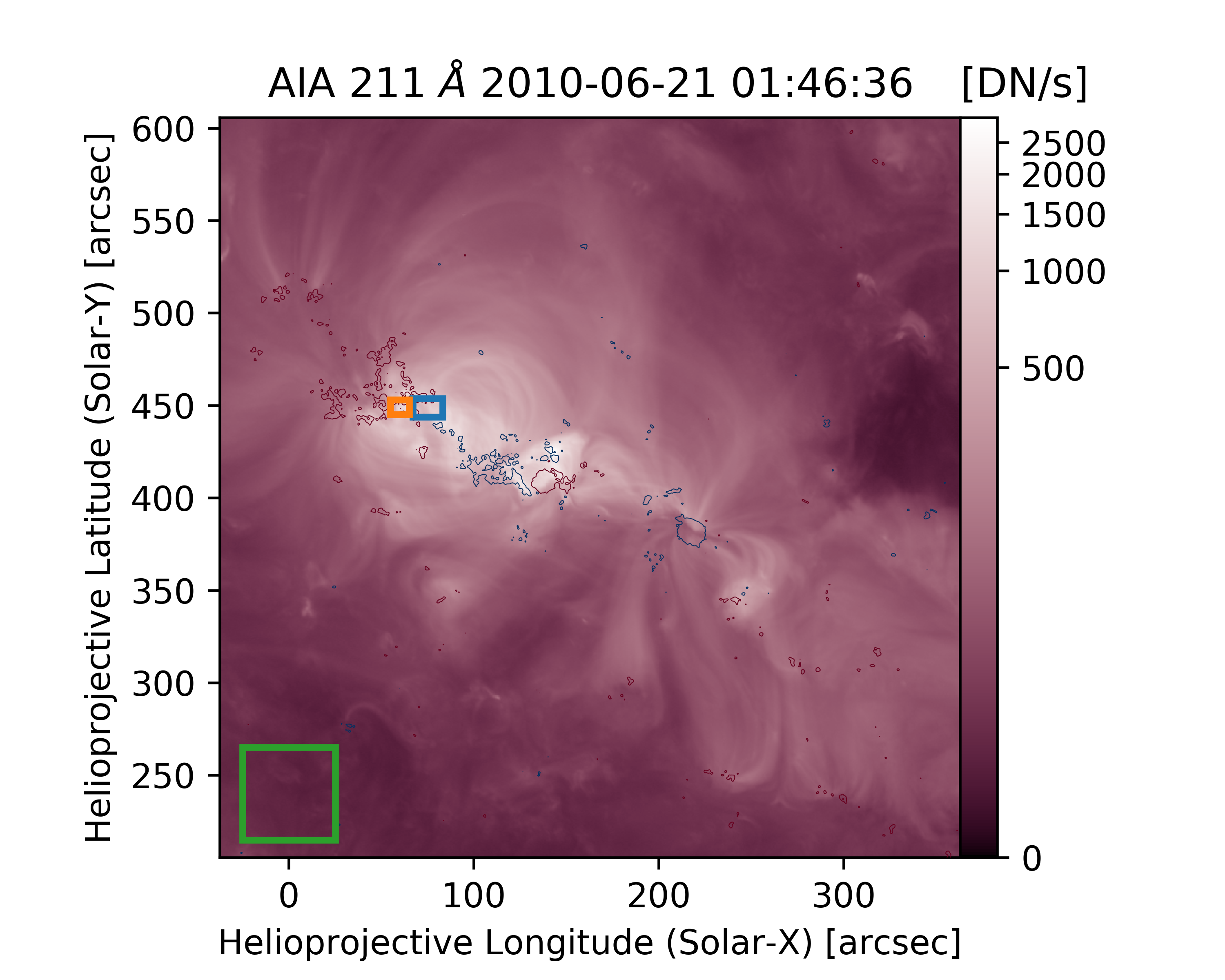}
	\includegraphics[trim=1.225cm 0.55cm 1.25cm 0.5cm, clip, width = 0.24\textwidth]
{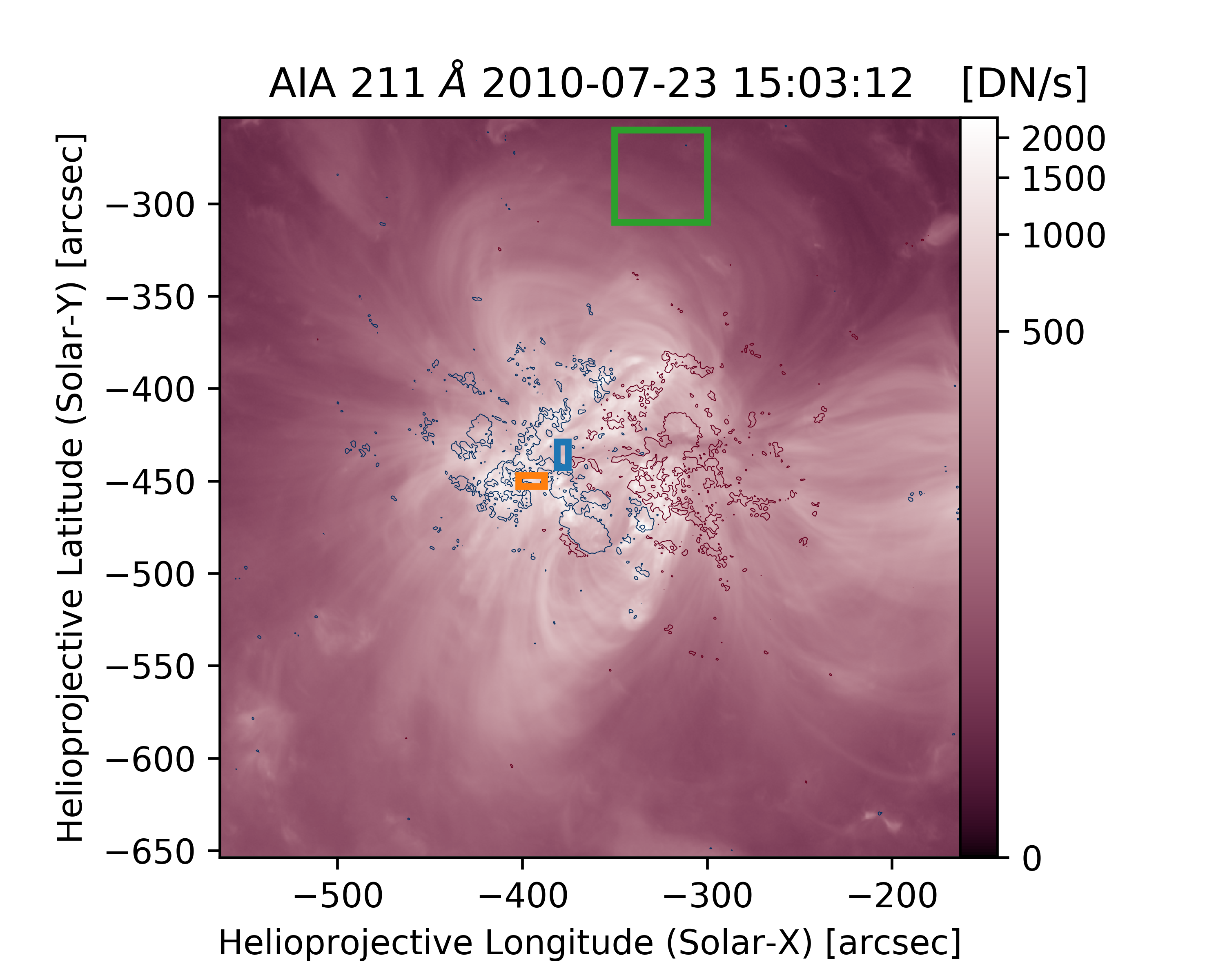}
	\includegraphics[trim=1.225cm 0.55cm 1.25cm 0.5cm, clip, width = 0.24\textwidth]
{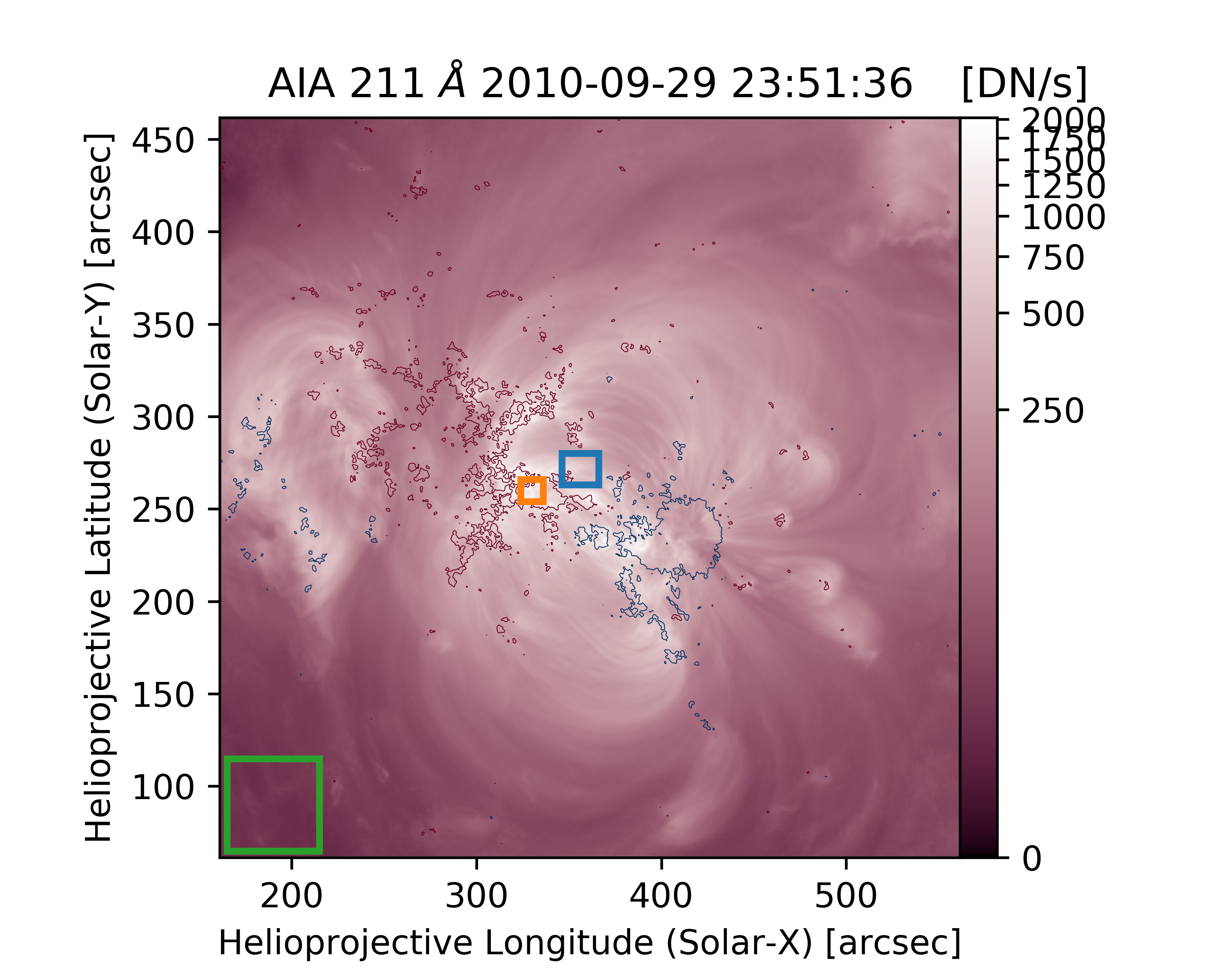}}
	\vspace{-0.5cm}
	\gridline{
	\includegraphics[trim=0.75cm 1.4cm 11.5cm 1.4cm, clip, width = 0.0125\textwidth]
{2011-08-09_211A.png}  % Left axis title
	\includegraphics[trim=1.21cm 0.55cm 1.25cm 0.5cm, clip, width = 0.24\textwidth]
{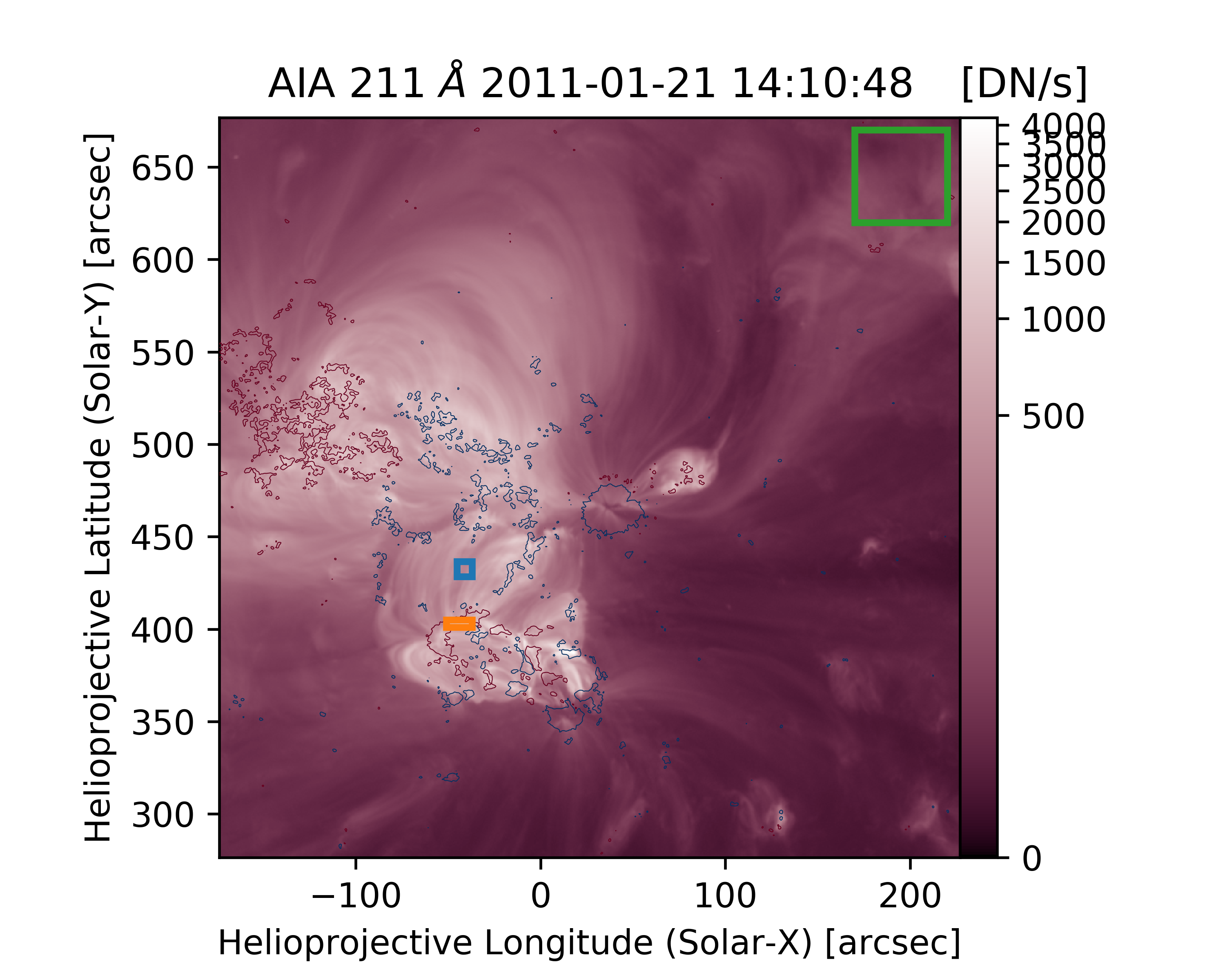}
	\includegraphics[trim=1.21cm 0.55cm 1.25cm 0.5cm, clip, width = 0.24\textwidth]
{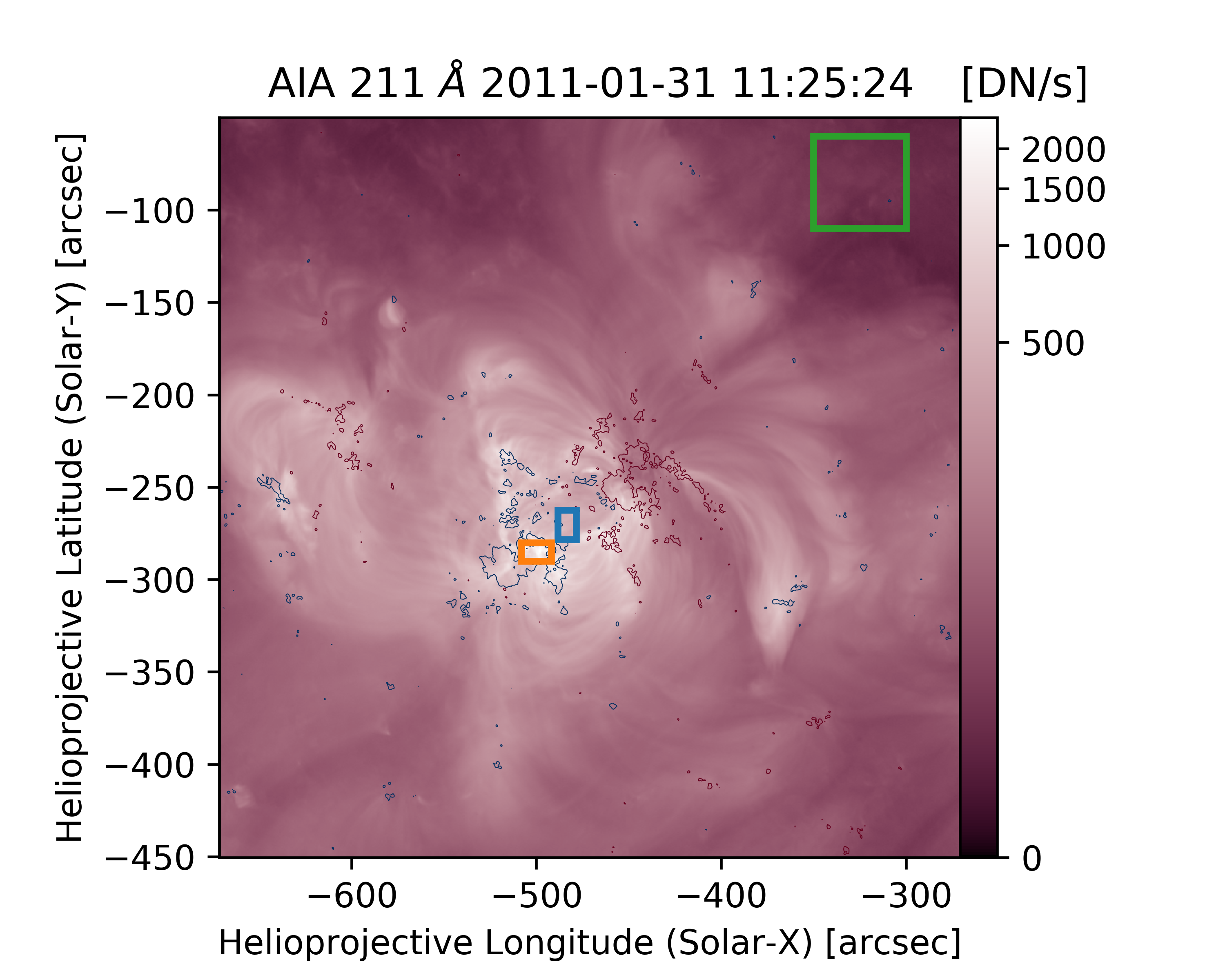}
	\includegraphics[trim=1.21cm 0.55cm 1.25cm 0.5cm, clip, width = 0.24\textwidth]
{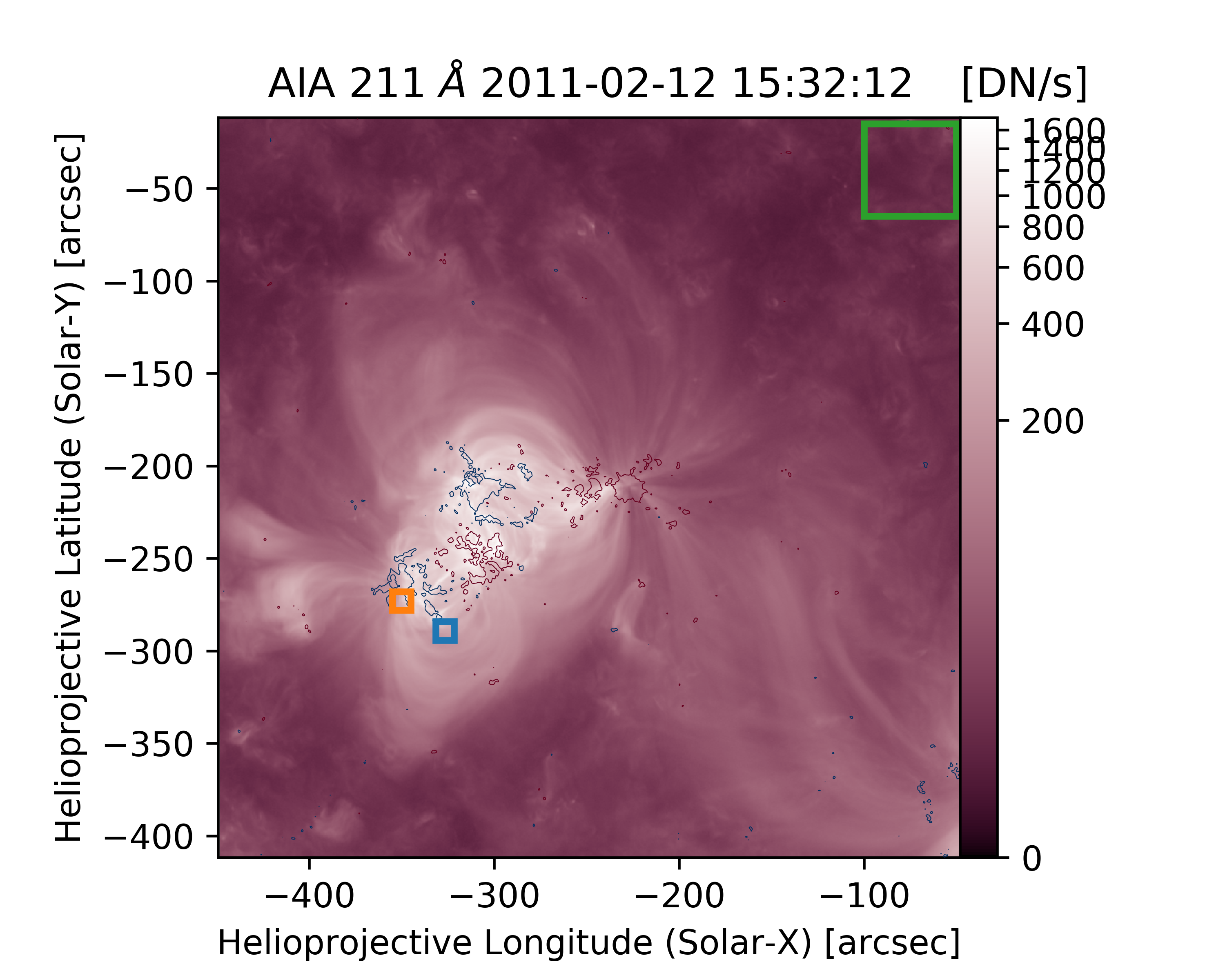}
	\includegraphics[trim=1.21cm 0.55cm 1.25cm 0.5cm, clip, width = 0.24\textwidth]
{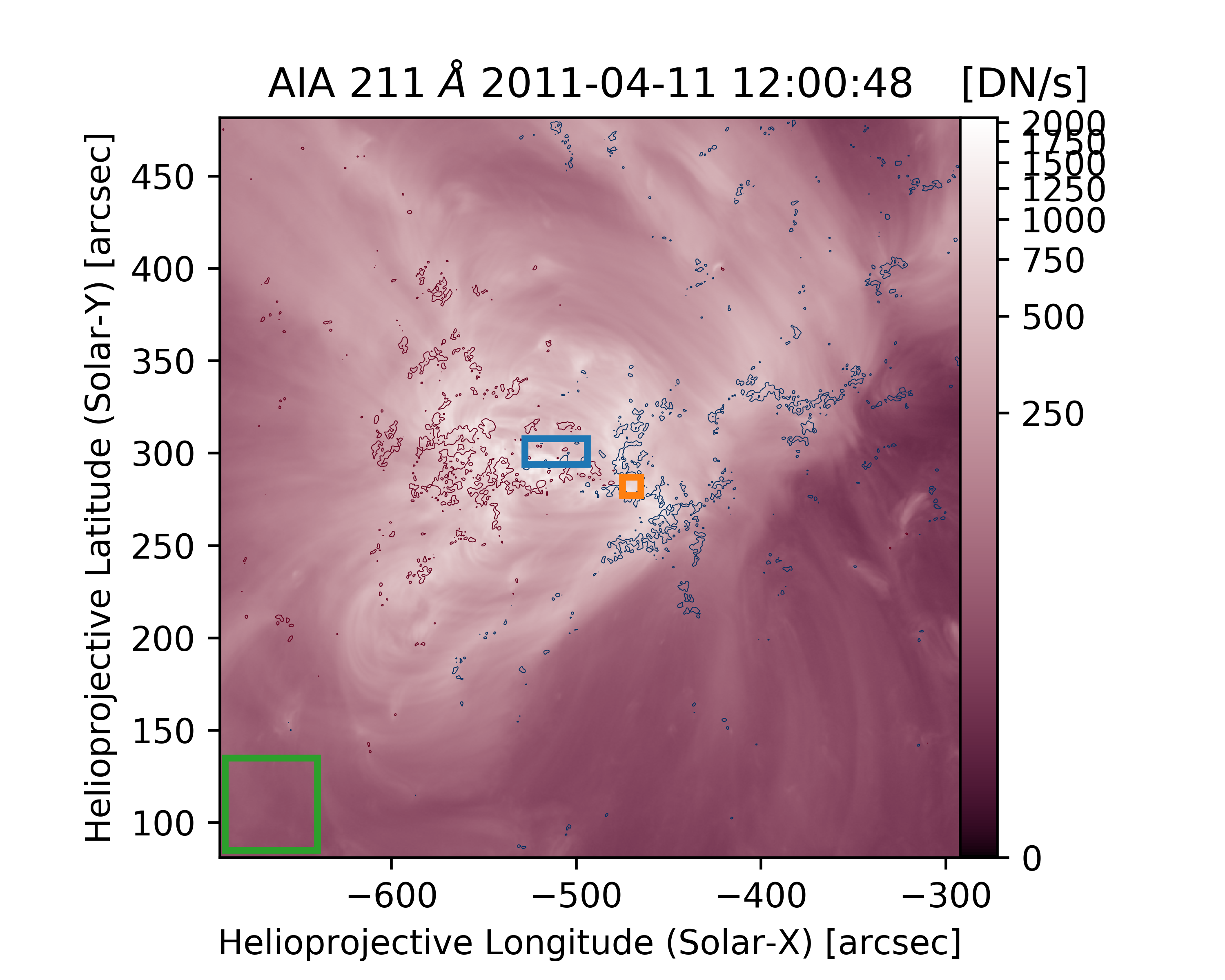}}
	\vspace{-0.5cm}
	\gridline{
	\includegraphics[trim=0.75cm 1.4cm 11.5cm 1.4cm, clip, width = 0.0125\textwidth]
{2011-08-09_211A.png}  % Left axis title
	\includegraphics[trim=1.225cm 0.55cm 1.25cm 0.5cm, clip, width = 0.24\textwidth]
{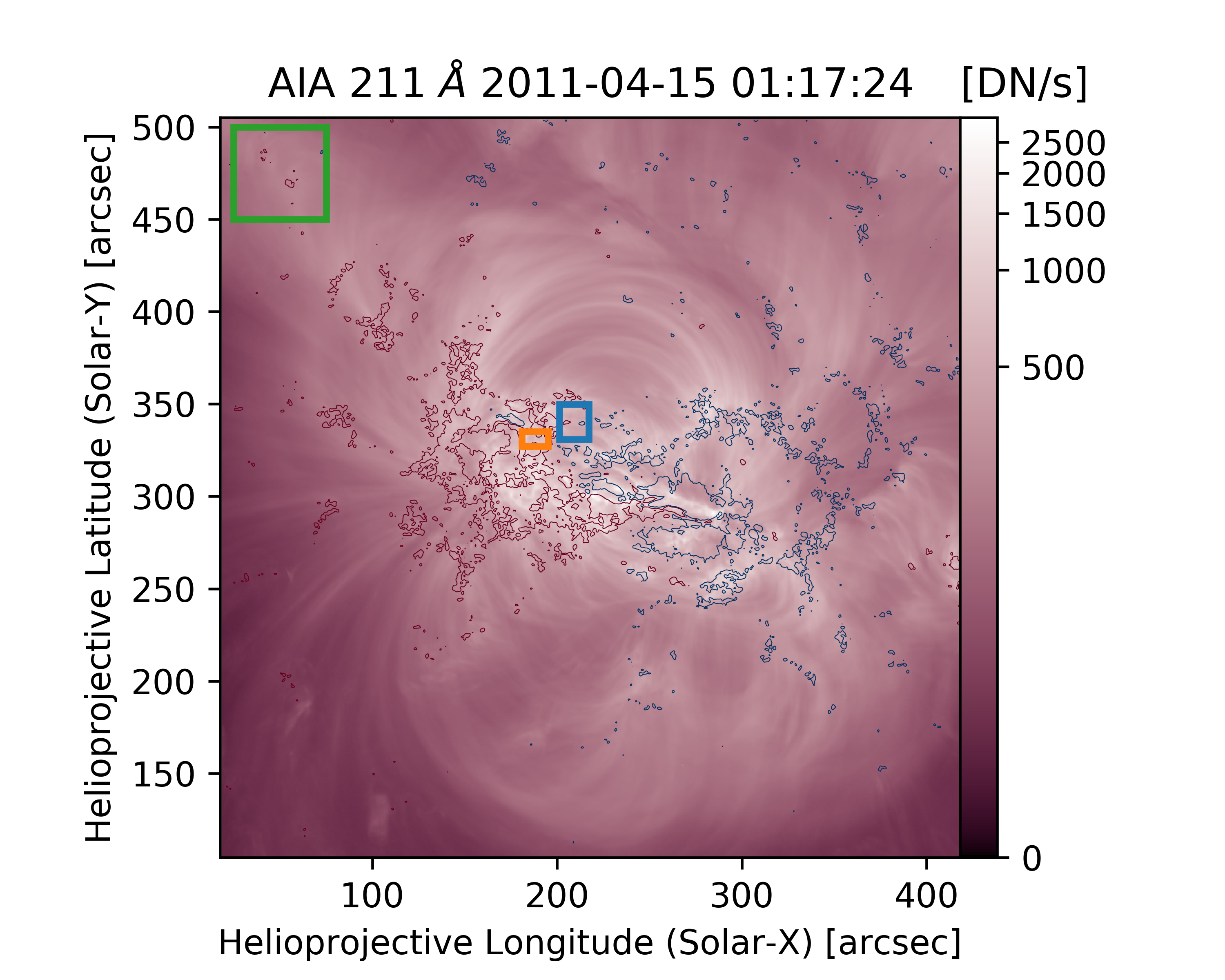}
	\includegraphics[trim=1.225cm 0.55cm 1.25cm 0.5cm, clip, width = 0.24\textwidth]
{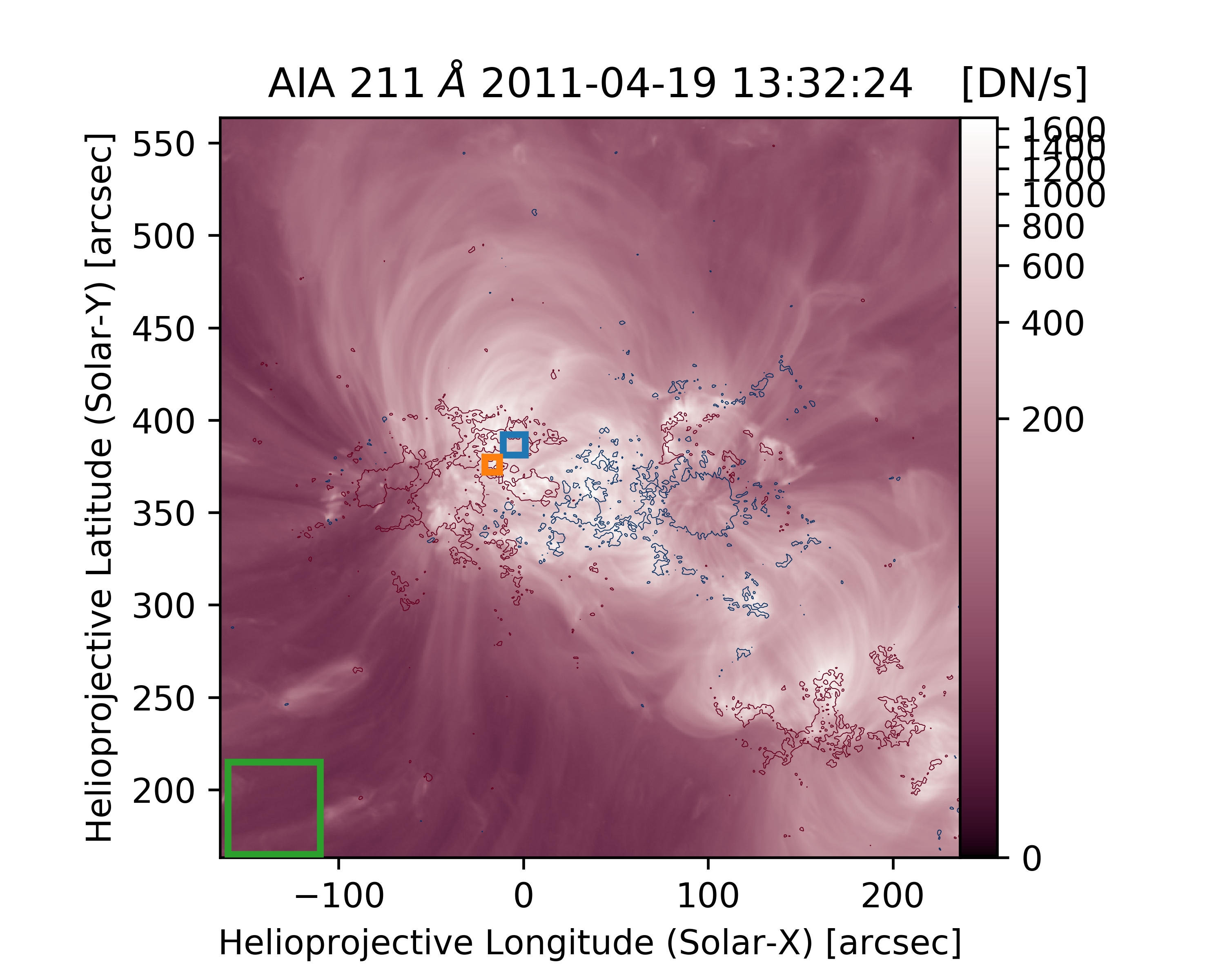}
	\includegraphics[trim=1.225cm 0.55cm 1.25cm 0.5cm, clip, width = 0.24\textwidth]
{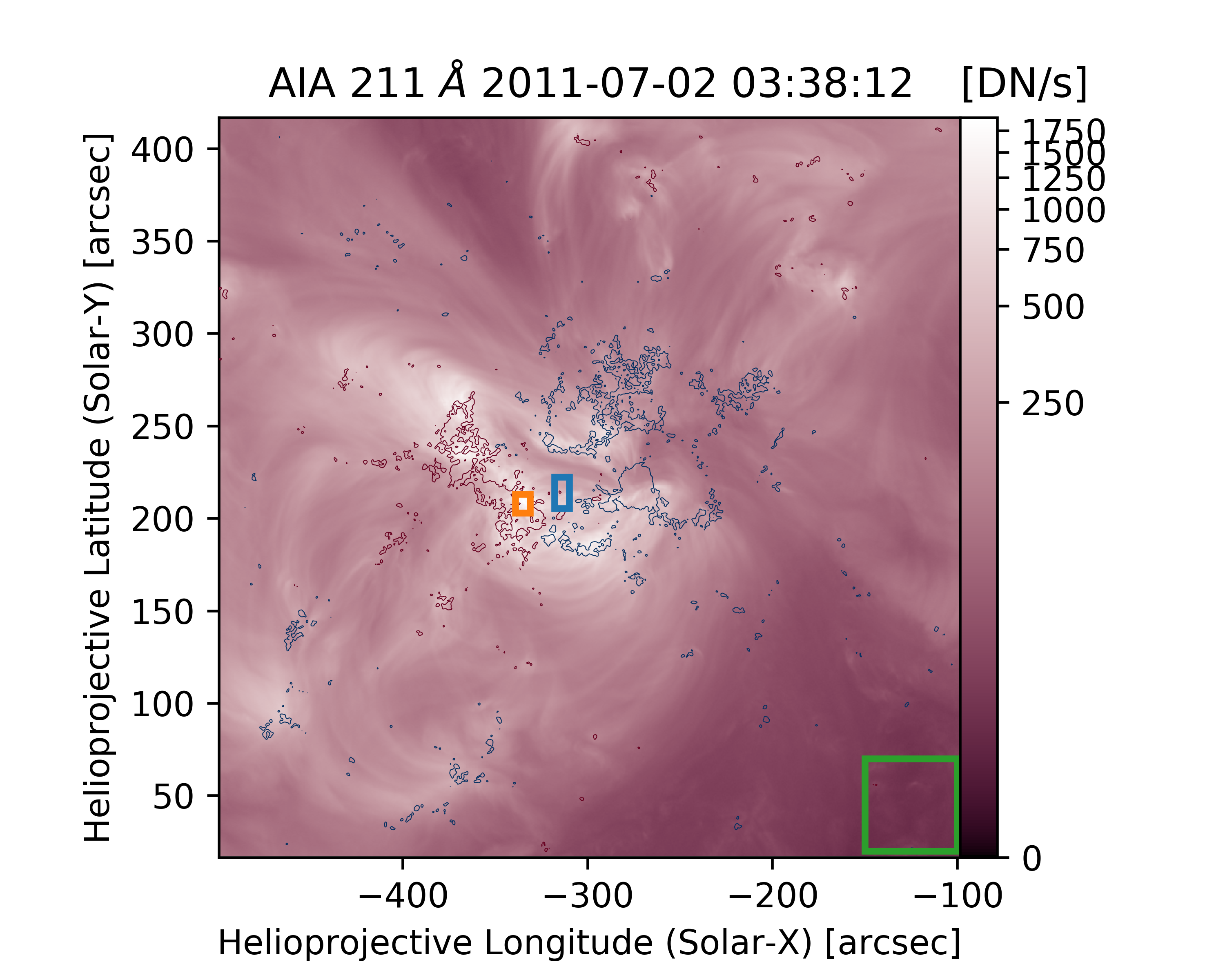}
	\includegraphics[trim=1.225cm 0.55cm 1.25cm 0.5cm, clip, width = 0.24\textwidth]
{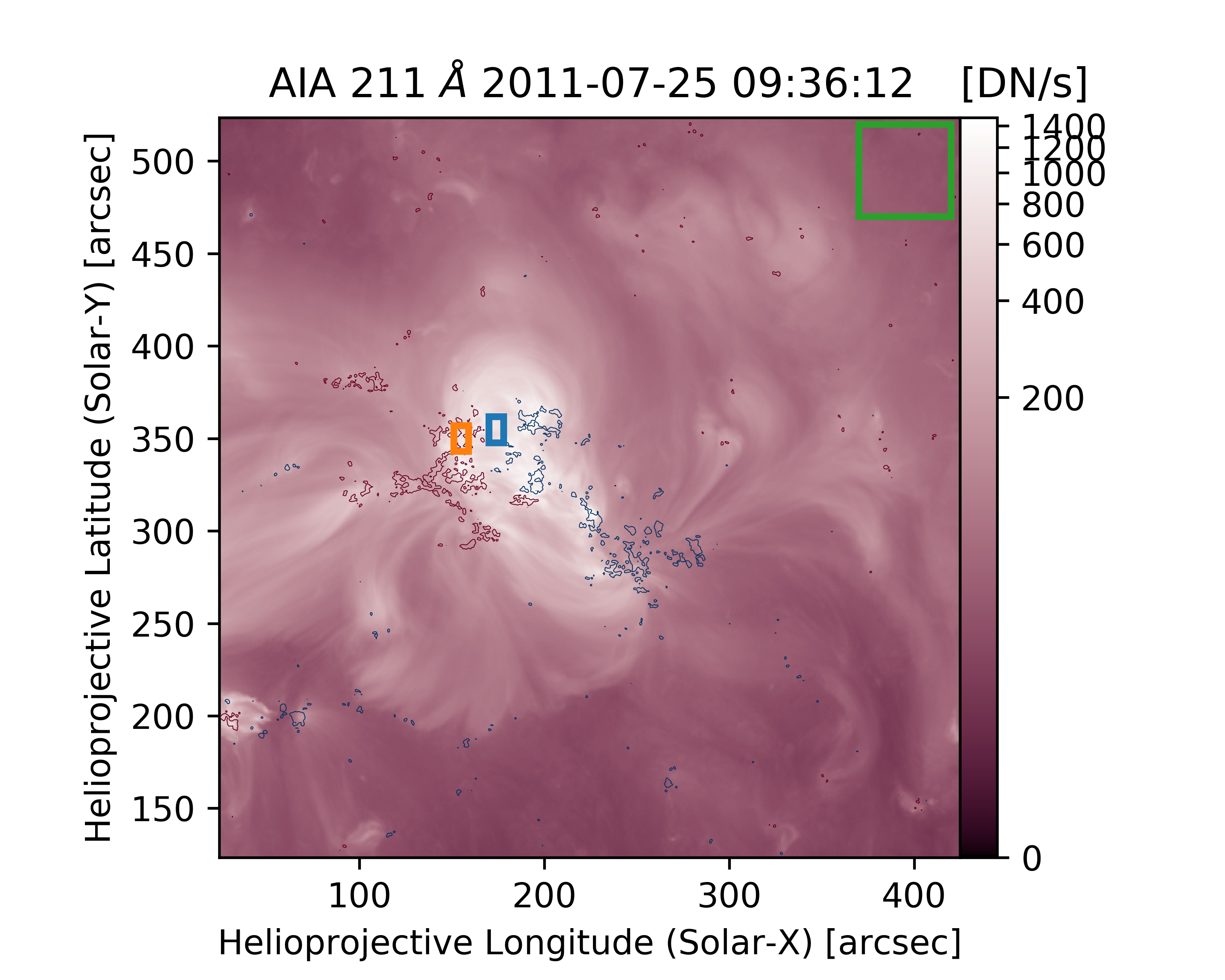}}
	\vspace{-0.5cm}
	\gridline{
	\includegraphics[trim=0.75cm 1.4cm 11.5cm 1.4cm, clip, width = 0.0125\textwidth]
{2011-08-09_211A.png}  % Left axis title
	\includegraphics[trim=1.225cm 0.55cm 1.25cm 0.5cm, clip, width = 0.24\textwidth]
{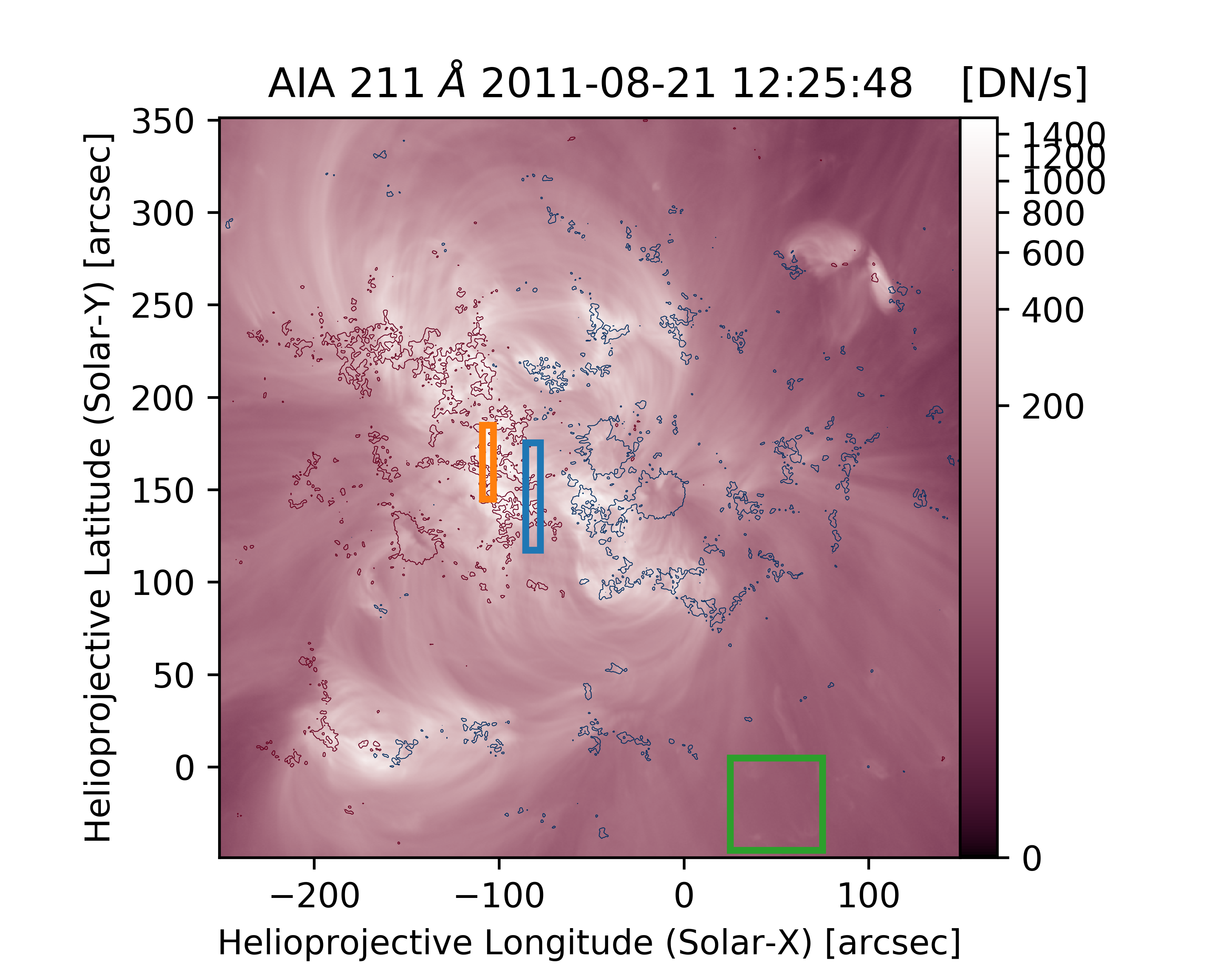}
	\includegraphics[trim=1.225cm 0.55cm 1.25cm 0.5cm, clip, width = 0.24\textwidth]
{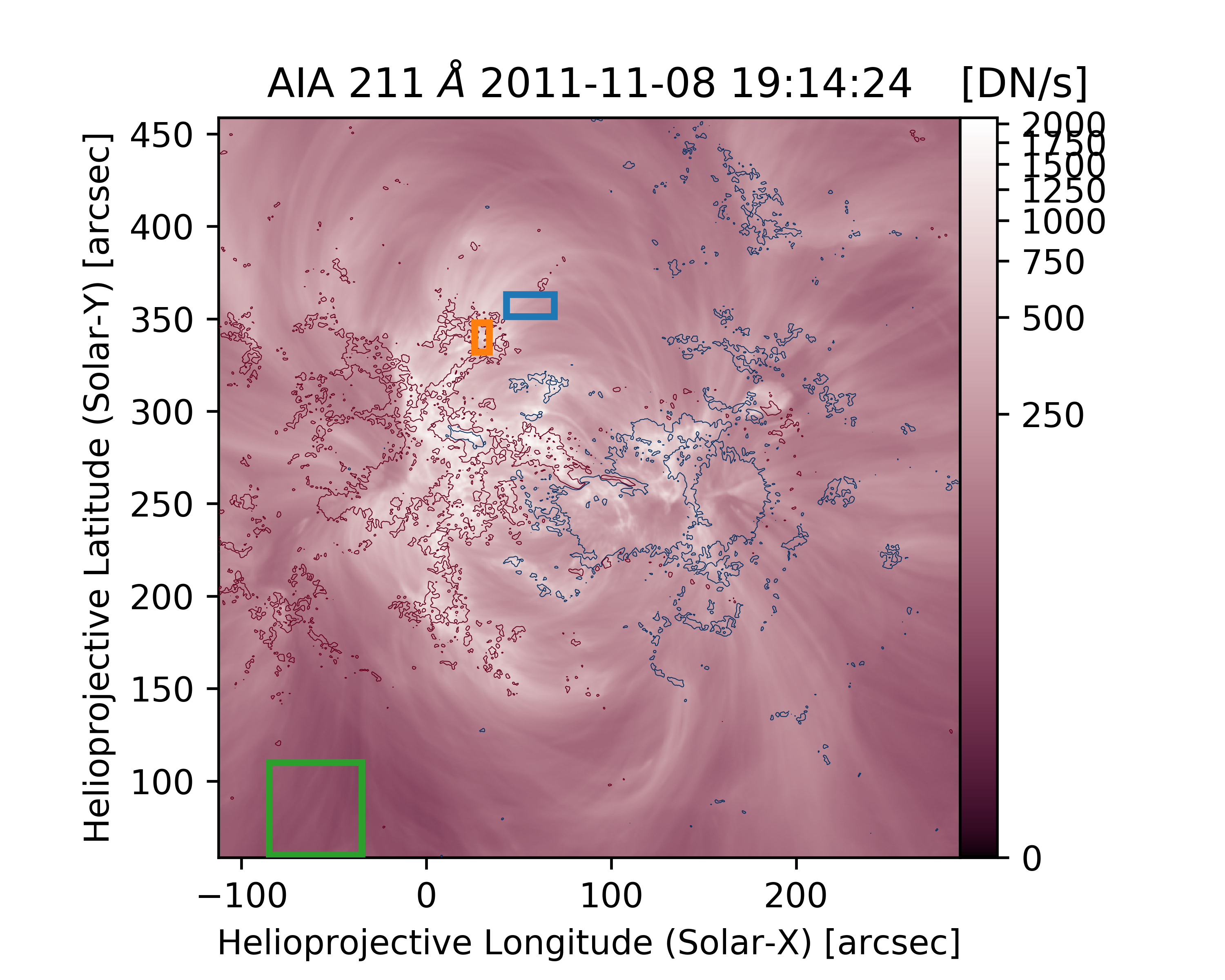}
	\includegraphics[trim=1.225cm 0.55cm 1.25cm 0.5cm, clip, width = 0.24\textwidth]
{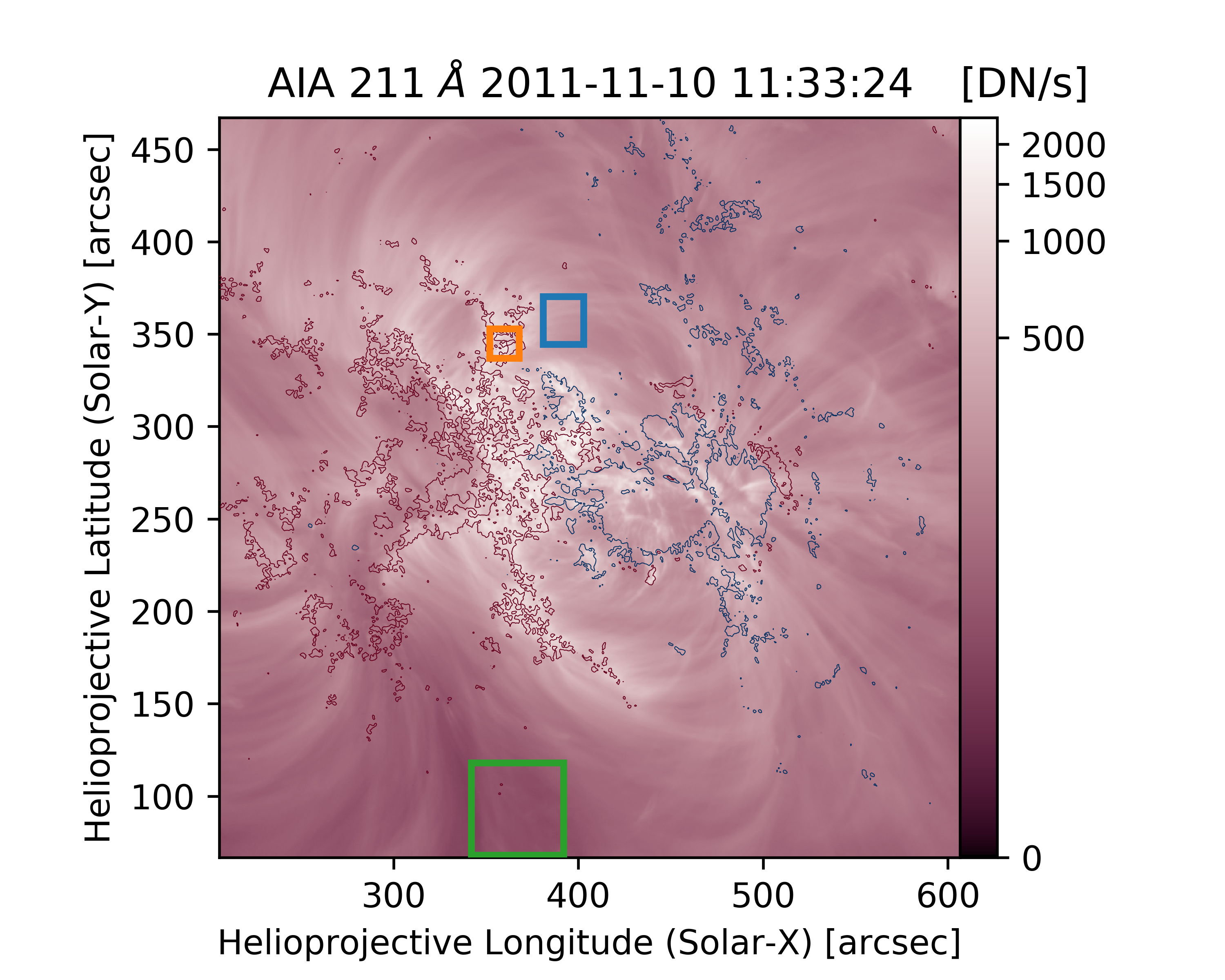}
	\includegraphics[trim=1.225cm 0.55cm 1.25cm 0.5cm, clip, width = 0.24\textwidth]
{2011-08-09_211A.png}}
	\vspace{-0.5cm}
	\gridline{
	\includegraphics[trim=0.75cm 0.2cm 11.5cm 9.55cm, clip, width = 0.012\textwidth]
{2011-08-09_211A.png}  % Left axis title fudge
	\includegraphics[trim=2.25cm 0.2cm 2.5cm 9.55cm, clip, width = 0.24\textwidth]
{2011-08-09_211A.png}  % Bottom axis title
	\includegraphics[trim=2.25cm 0.2cm 2.5cm 9.55cm, clip, width = 0.24\textwidth]
{2011-08-09_211A.png}  % Bottom axis title
	\includegraphics[trim=2.25cm 0.2cm 2.5cm 9.55cm, clip, width = 0.24\textwidth]
{2011-08-09_211A.png}  % Bottom axis title
	\includegraphics[trim=2.25cm 0.2cm 2.5cm 9.55cm, clip, width = 0.24\textwidth]
{2011-08-09_211A.png}}  % Bottom axis title
	\caption{AIA $211$ \AA\ five minute average images of the active regions from \cite{Warren2012}. The boxed regions and contours highlight the same features as in Figure \ref{fig:aia:region}. The \textit{bottom right} panel shows the individual region from Figure \ref{fig:aia:region} for comparison.}
	\label{fig:aia:regions}
\end{figure*}

For each of the 15 active regions studied in \cite{Warren2012} (except their Region 13 Box 2) we repeat the analysis performed in Section \ref{sec:observations:separating}. The regions identified as coronal are those defined and analyzed in the original paper while the transition region boxes are determined by eye based on the apparent connectivity of the loop features in each region. These active regions and the associated boxes indicating the corona, transition region, and quiet Sun are shown in Figure \ref{fig:aia:regions}. We expect that the wide range of active region structures and viewing geometries represented in this sample will minimize any particular geometrical bias introduced by analyzing a single region. In addition, these regions represent a wide range of physical scales with potentially different heating properties.

\begin{figure}[t]
	\includegraphics*[trim= 0.1cm 0.25cm 0.25cm 0.25cm, clip, width=\columnwidth]
{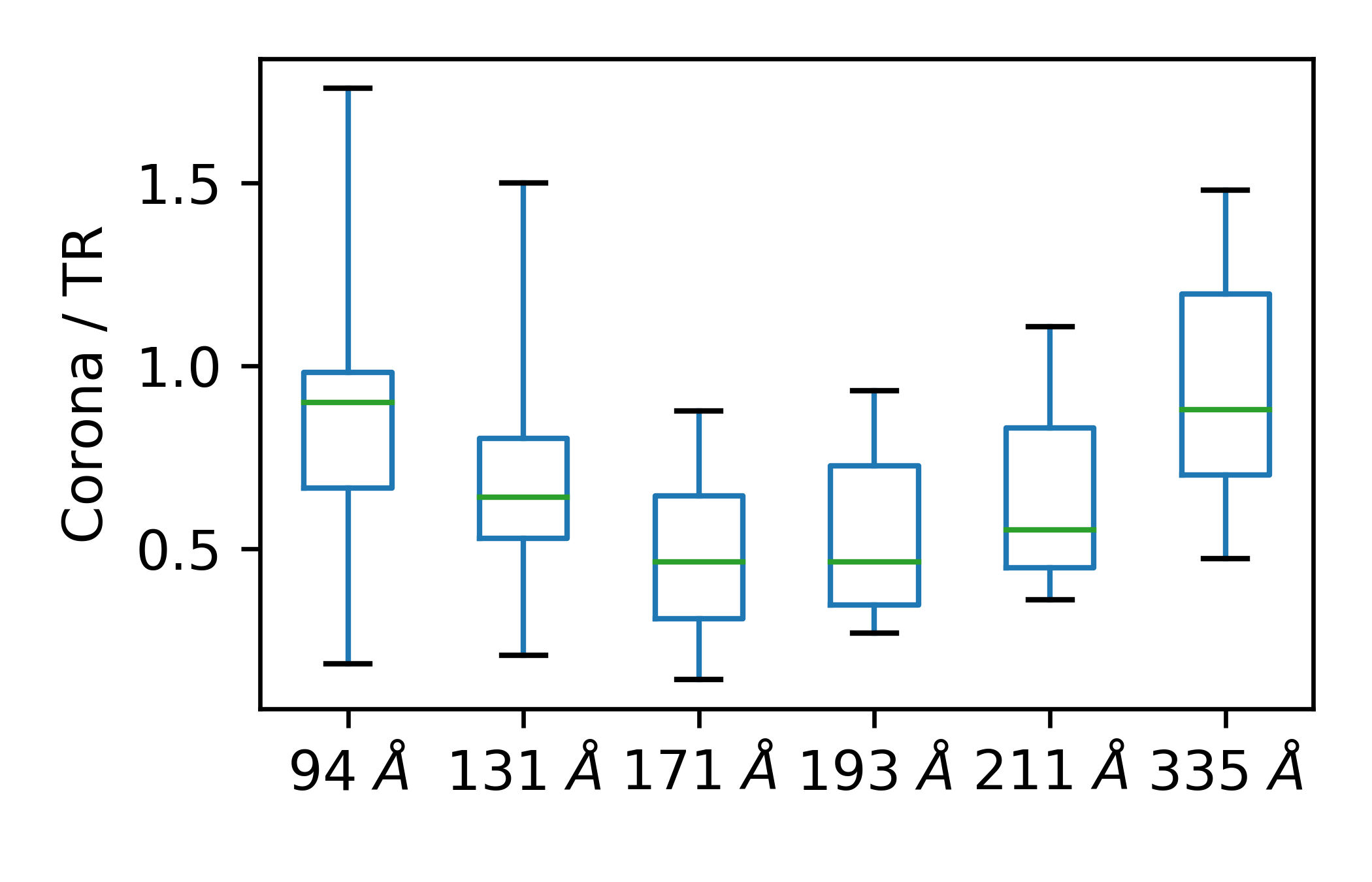}
	\caption{Observed R$_{\text{C/TR}}$ in the \cite{Warren2012} active regions. The green lines indicate the median of all the active regions, the box indicates the lower and upper quartiles, and the whiskers indicate the extremes. This plot is equivalent to the blue bars in Figure \ref{fig:aia:region:ratios}}
	\label{fig:aia:regions:ratios}
\end{figure}

We compute R$_{\text{C/TR}}$ in each channel for each active region individually. The distributions of these ratios are plotted in Figure \ref{fig:aia:regions:ratios}. Notice that while the ratios are on average larger than the ratios found in NOAA 11268, in most cases the transition region is still brighter than the corona. Only in the $94$ \AA\ and $335$ \AA\ channels is this not generally the case. Previous analysis of these active regions determined that they have DEMs peaking between log(T [K]) = $6.5$--$6.6$, where the $94$ \AA\ and $335$ \AA\ channels have the highest relative response. It is not surprising, therefore, that R$_{\text{C/TR}}$ is greatest in these channels. While there is some plasma above the DEM peak, the slopes of the DEMs are quite steep, and there is very little plasma at log(T [K]) $\sim7.1$, the temperature of the strong secondary peak in the $131$ \AA\ channel. The $131$ \AA\ intensity ratios are consequently smaller, although still elevated compared to NOAA 11268.

\begin{figure}[t]
	\includegraphics[trim=0.4cm 0.4cm 0.35cm 0.35cm, clip, width=\columnwidth]
{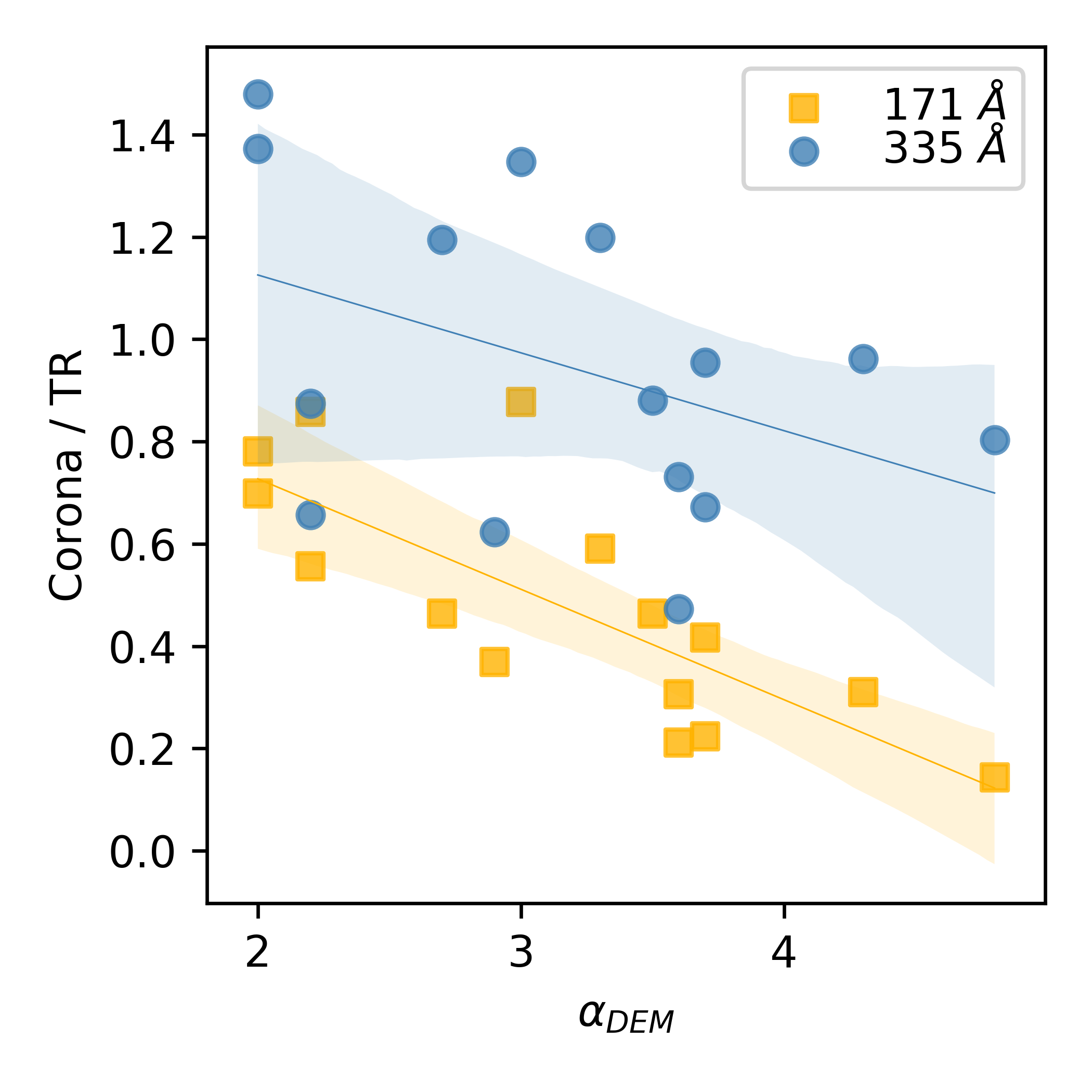}
	\caption{Correlation between R$_{\text{C/TR}}$ and $\alpha_{\text{DEM}}$ identified in \cite{Warren2012} for each active region in the $171$ \AA\ and $335$ \AA\ channels. The best-fit linear relationship and error region for each channel are plotted to guide the eye. Statistics about the relationship between R$_{\text{C/TR}}$ and $\alpha_{\text{DEM}}$ for each channel are given in table \ref{tab:correlations}.}
	\label{fig:aia:regions:correlations}
\end{figure}

\begin{deluxetable}{lcccc}
  \tablewidth{0pt}
  \tablecaption{Correlation between $\alpha_{\text{DEM}}$ and R$_{\text{C/TR}}$}
  \tablecolumns{5}
  \tablehead{AIA channel & \colhead{r} &
  	\colhead{P(tw)} & \colhead{$\chi$} &
  	\colhead{$\chi_{90\%}$}}
  \startdata
   $\;\;94$ \AA & $-0.21$ & $0.678$ & $-0.10$ & $-1.06 :\;\;\;0.32$ \\  % \; add spaces for alignment
  	$131$ \AA & $-0.45$ & $0.026$ & $-0.74$ & $-1.78 : -0.26$ \\
  	$171$ \AA & $-0.78$ & $0.004$ & $-1.78$ & $-2.42 : -1.04$ \\
  	$193$ \AA & $-0.73$ & $0.010$ & $-1.12$ & $-1.82 : -0.48$ \\
  	$211$ \AA & $-0.68$ & $0.040$ & $-0.80$ & $-1.37 : -0.41$ \\
  	$335$ \AA & $-0.42$ & $0.537$ & $-0.50$ & $-1.16 :\;\;\;0.32$ \\
  \enddata
  \tablecomments{r is the Pearson correlation coefficient between $\alpha_{\text{DEM}}$ and R$_{\text{C/TR}}$. P(tw) is the probability of drawing the observed distribution from a uniform random distribution. $\chi$ is the most probable exponential in the relationship $\text{R}_\text{C/TR}\propto\left(\alpha_{\text{DEM}}\right)^{\chi}$. $\chi_{90\%}$ is the $90\%$ confidence interval of $\chi$.}
  \label{tab:correlations}
\end{deluxetable}

\cite{Warren2012} measured the power-law index of the DEM distribution in the range $6.0\ \leq$ log(T) $\leq\ 6.6$ (approximately the peak temperature) in each of the coronal boxes in Figure \ref{fig:aia:regions}. This is the slope, $\alpha_{\text{DEM}}$, in a log-log plot. We compare the coronal DEM slopes with R$_{\text{C/TR}}$. A sample of these relationships is shown for the $171$ \AA\ and $335$ \AA\ channels in Figure \ref{fig:aia:regions:correlations}. There is a clear anticorrelation in the $171$ \AA\ channel, in which larger intensity ratios correspond to smaller slopes, i.e., flatter DEM distributions. The same trend appears in the $335$ \AA\ channel, but with much larger scatter. To quantify the trends, we perform multiple statistical analyses, as reported in table \ref{tab:correlations}. Because the distributions appear approximately linear, we compute the Pearson correlation coefficient. The negative coefficients indicate the inverse relationships while the larger magnitudes of the $171$ \AA, $193$ \AA, and $211$ \AA\ channels indicate tighter correlations (less scatter).

One disadvantage of the Pearson analysis is that it assumes that the measured quantities are normally distributed, i.e., that the errors in the measurements follow a normal distribution. We have no indication that this is or is not the case. We therefore also perform a nonparametric, or rank ordered, statistical analysis, which is valid for any measurement distribution. We use the weighted t-statistic described in \cite{Efron1992}, following the implementation in \cite{Porter1995}. The probability that $\alpha_{\text{DEM}}$ and R$_{\text{C/TR}}$ are random is given by P(tw) in table \ref{tab:correlations}. A small value indicates a high probability of correlation. The fourth column indicates the most probable $\chi$ in the assumed relationship $\text{R}_\text{C/TR}\propto\left(\alpha_{\text{DEM}}\right)^{\chi}$, and the final column gives the $90\%$ confidence interval of $\chi$. From these analyses, we see that all channels except $94$ \AA\ and $335$ \AA\ have robust inverse correlations. The relationship between $\alpha_{\text{DEM}}$ and R$_{\text{C/TR}}$ in the $94$ \AA\ and $335$ \AA\ channels is likely random, which could be due to their significantly nonisothermal temperature response functions. Again, the $131$ \AA\ channel is functionally isothermal in these observations because there is very little plasma above $10$ MK in these regions.

\cite{Warren2012} also measured the slopes of the coronal DEM with log(T) $\geq 6.6$, hotter than the peak. We compare those slopes with the intensity ratios and find no significant correlation in any channel.

We can offer a partial explanation for the robust inverse correlation between $\alpha_{\text{DEM}}$ and R$_{\text{C/TR}}$ in the $131$ \AA, $171$ \AA, $193$ \AA, and $211$ \AA\ channels. Consider, for example, the $211$ \AA\ channel with a peak response at $2$ MK. This channel measures the corona of loops with coronal temperatures near $2$ MK but the transition region of loops with coronal temperatures near $4$ MK. Fundamentally, R$_{\text{C/TR}}$ in a given channel correlates positively with emission measure at the peak of the temperature response function and negatively with emission measure at temperatures greater than about twice the peak of the temperature response function. However, the exact explanation depends on the frequency with which the plasma is heated.

%Fundamentally, R$_{\text{C/TR}}$ correlates positively with plasma at the peak of the temperature response function and negatively with plasma at temperatures greater than twice the peak of the temperature response function. These channels are sensitive to the transition region emission in the hot ($4$ MK) strands, but not in significantly cooler strands. But 

\begin{itemize}
\item In the case of high-frequency heating, individual strands evolve very little. A shallow coronal DEM slope (small $\alpha_{\text{DEM}}$) indicates that nearly as many strands are held at a quasi-constant coronal temperature of, say, $2$ MK as are held at a quasi-constant coronal temperature of $4$ MK. A steep slope (large $\alpha$) indicates the dominance of hot strands. Consequently, R$_{\text{C/TR}}$ will be smaller (relatively brighter transition region) when the slope is steep (relatively more hot strands).

\item For low-frequency heating, the same argument as discussed in Section \ref{sec:simulation:AIA} applies. Strands experiencing low-frequency heating that begin their cooling from higher initial temperatures have smaller R$_{\text{C/TR}}$. Because $\alpha_{\text{DEM}}$ is calculated over a fixed temperature range ($6.0\ \leq$ log(T) $\leq\ 6.6$), relatively more strands heated to peak temperatures coolward of log(T) = $6.6$ flatten the DEM (decreasing $\alpha_{\text{DEM}}$) and result in larger R$_{\text{C/TR}}$.

\item In the intermediate frequency heating regime, strands cool partially before being reheated. A steep DEM slope (large $\alpha_{\text{DEM}}$) indicates that relatively more strands begin their cooling at a higher maximum temperature and/or are reheated before cooling to lower temperatures. The same arguments that explain the anticorrelation between R$_{\text{C/TR}}$ and $\alpha_{\text{DEM}}$ in the low-frequency heating case apply here, with an additional, reinforcing effect. If the coronal segment of a strand never cools through the peak response of a given channel, that channel will collect even less coronal emission leading to a smaller R$_{\text{C/TR}}$.
\end{itemize}

These effects are not uniform across all AIA channels and depend on the shape of the temperature response function, but apply generally to the $131$ \AA, $171$ \AA, $193$ \AA, and $211$ \AA\ channels that are quasi-isothermal with peak response below $2$ MK. We also note that no model in Section \ref{sec:simulation} has exclusively high-, intermediate-, or low-frequency heating. They all include a mixture of the three, with the relative proportions being different from model to model. The same is likely true in these observed active regions.

\section{Conclusion}
\label{sec:conclusion}
Using the computational efficiency of EBTEL modeling and active regions studied by \cite{Warren2012} we investigated the theoretical and observed contribution of the transition region to AIA images. For this analysis, we defined the transition region from a physically meaningful perspective as the volume of the solar atmosphere above the chromosphere that is heated (while the corona is cooled) by thermal conduction, rather than more traditional observational definitions based on plasma temperature. With this definition, the transition region is confined to low altitudes, as in the conventional picture. This study involved two major investigations: an exploration of the parameter space of relevant coronal heating variables, with particular focus on the frequency of impulsive heating events, and a study of observed active regions to provide an observational anchor for the models.

The EBTEL models revealed that, consistent with previous studies \citep[e.g.;][]{Patsourakos2008}, imaging observations often described as ``coronal'' are expected to have significant contribution from transition region plasma. We find that the ratio of coronal to transition region emission is very different for the individual AIA channels and depends strongly on the heating parameters, demonstrating promising diagnostic potential. In general, we find that those scenarios with higher frequency heating events lead to higher time-averaged coronal temperatures and densities, but lower maximum temperatures and densities. However, observed intensities depend on the full DEM distribution, including both the coronal and transition region contributions, and it is not possible to easily predict the brightness in a channel based on the time-averaged coronal temperature and density alone. We also find that those strands subjected to the highest frequency heating agree quite well with theoretical expectations for coronal loops in static equilibrium. Overall, our analysis suggests that in shorter strands, the emission from the transition region and corona are comparable, while the emission from long strands tends to be dominated by the transition region, particularly in the higher frequency heating scenarios.

We performed a simple analysis of observed AIA active regions, comparing the intensity of emission from coronal and transition region plasma identified based on their morphology and relation to photospheric magnetic fields. Analyzing observations of active region NOAA 11268, we find an overall consistency with the models. The observations confirm the general trend in the models that the $335$ \AA, $211$ \AA, and sometimes $94$ \AA\ channels (i.e. those associated with the hotter plasma) have the largest ratios and the $131$ \AA, $171$ \AA, and $193$ \AA\ channels have the smallest ratios. The observed ratios depend, however, on assumptions about how much overlying coronal emission is present above the footpoint transition region emission. These same observational trends persist when analyzing the 15 active regions from \cite{Warren2012}, although they have generally higher ratios. All of these active regions suggest that AIA observations of loops sample a similar level of emission from the corona and transition region.

We also analyzed the relationships between R$_{\text{C/TR}}$ and the slopes of the DEMs determined by \cite{Warren2012}. We find that there is a consistent negative relationship between the slope of the DEM coolward of the temperature peak and R$_{\text{C/TR}}$ in the observed regions. This is consistent with theoretical expectations based on low, intermediate, or high frequency impulsive heating.

We note that, particularly for the longer $80$ Mm strands, the models suggest that the ratio of coronal to transition region intensity should be significantly smaller than is observed. One potential explanation for this is the absorption of transition region emission from spicules extending from the underlying chromosphere. This has been found to cause up to a factor of 2 decrease in the observed transition region intensity \citep{DePontieu2009}, which would increase the observed ratios compared to model predictions, consistent with our findings.

We made no attempt to ascribe a particular heating model to the studied active regions because individual zero-dimensional EBTEL models are inadequate to properly characterize the complexity of active region observations. It is unreasonable to expect the model of a single magnetic strand to replicate observations from even simple active regions. In addition, there is ambiguity due to the somewhat arbitrary choice of observational path length assigned to the coronal emission in the EBTEL models. Both of these uncertainties can be largely resolved by studying this effect in three-dimensional models of active regions where the true extent of the corona can be more accurately estimated. We have begun to construct such models, based on observed photospheric magnetograms, using the approach described in \cite{Nita2018}.

Despite the idealized nature of the modeling and observational analyses presented here, they clearly demonstrate the importance of considering the transition region in active region models, particularly when they are used to study coronal heating. Depending on how the active region is heated, failing to include the transition region could lead to significant underestimation of the AIA emission from the region.

\acknowledgments
Data supplied courtesy of the SDO/HMI and SDO/AIA consortia. SDO is the first mission launched for NASA's Living With a Star (LWS) Program. EBTEL++ is developed and maintained by the Rice University Solar Physics Research Group. The authors would like to thank Harry Warren for providing data from \cite{Warren2012}. SJS's research was supported by an appointment to the NASA Postdoctoral Program at the Goddard Space Flight Center, administered by Universities Space Research Association under contract with NASA. This work of JAK was supported by the Goddard Space Flight Center Internal Scientist Funding Model (competitive work package) program.

\bibliography{library}

\end{document}